\documentclass[apj]{emulateapj}
\usepackage{multirow}
\usepackage{booktabs}
\usepackage{graphicx}
\usepackage{amsmath}
\usepackage{longtable}
\usepackage{rotating}
\usepackage{enumerate}
\usepackage{hyperref}

\newcommand{\mj}{M_{\rm J}}

\newcommand{\rsun}{R_\odot}
\newcommand{\msun}{M_\odot}
\newcommand{\re}{R_\oplus}
\newcommand{\me}{M_\oplus}
\newcommand{\rplanet}{R_{\rm p}}
\newcommand{\mplanet}{M_{\rm p}}
\newcommand{\rp}{r_{\rm p}}
\newcommand{\mth}{M_{\rm thermal}}
\newcommand{\mstar}{M_\star}
\newcommand{\sigmag}{\Sigma_{\rm gas}}
\newcommand{\sigmad}{\Sigma_{\rm dust}}
\newcommand{\sigmasd}{\Sigma_{\rm small-dust}}
\newcommand{\sigmabd}{\Sigma_{\rm big-dust}}
\newcommand{\hg}{h_{\rm gas}}
\newcommand{\hsd}{h_{\rm small-dust}}
\newcommand{\hbd}{h_{\rm big-dust}}

\begin{document}
\title{Multiple Disk Gaps and Rings Generated by a Single Super-Earth}

\shorttitle{One Super-Earth, Many Gaps and Rings}

\shortauthors{Dong, Li, Chiang, \& Li}

\author{Ruobing Dong\altaffilmark{1,2}, Shengtai Li\altaffilmark{3}, Eugene Chiang\altaffilmark{4}, \& Hui Li\altaffilmark{3}}

\altaffiltext{1}{Steward Observatory, University of Arizona, Tucson, AZ, 85719; rdong@email.arizona.edu}
\altaffiltext{2}{Bok Fellow}
\altaffiltext{3}{Theoretical Division, Los Alamos National Laboratory, Los Alamos, NM 87545}
\altaffiltext{4}{Department of Astronomy, University of California at Berkeley, Berkeley, CA 94720}

\clearpage

\begin{abstract}
We investigate the observational signatures of super-Earths (i.e., Earth-to-Neptune mass planets),
the most common type of exoplanet discovered to date,
in their natal disks of gas and dust.
Combining two-fluid global hydrodynamics simulations with
a radiative transfer code, we calculate the distributions of
gas and of sub-mm-sized dust in a disk perturbed by a super-Earth,
synthesizing images in near-infrared scattered light
and the mm-wave thermal continuum for direct comparison with observations.
In low viscosity gas ($\alpha\lesssim10^{-4}$), a super-Earth opens
two annular gaps to either side of its orbit by the action of Lindblad
torques. This double gap and its associated gas pressure
gradients
cause dust particles to be dragged by gas into three rings:
one ring sandwiched between the two gaps, and two rings
located at the gap edges farthest from the planet. Depending on system
parameters, 
additional
rings may manifest for a single planet.
A double gap located at tens of AUs from a host star in Taurus
can be detected in the dust continuum 
by the Atacama Large Millimeter Array (ALMA) 
at an angular resolution of
$\sim$$0\arcsec.03$ 
after two hours of integration.
Ring and gap features persist in a variety of background disk profiles,
last for thousands of orbits, and change their relative positions and
dimensions depending on the speed and direction of
planet migration. Candidate double gaps have
been observed by ALMA in systems like HL Tau (D5 and D6) and
TW Hya (at 37 and 43 AU);
we submit that
each double gap
is carved
by one super-Earth in nearly inviscid gas.

\end{abstract}



\section{Introduction}\label{sec:intro}

Practically
every
star in our
Galaxy
harbors planets
\citep[e.g.,][]{fressin13, burke15, dressing15charbonneau}.
The most common planets discovered so far are 
``super-Earths,'' having radii $\rplanet$ intermediate between those of
Earth and Neptune (for reference, Neptune's $\rplanet = 4\re$ and its
mass $M_{\rm p} = 17 \me$).\footnote{Planets at
the upper end of this size range, $2\re\leq\rplanet\leq4\re$, are sometimes
called ``sub-Neptunes'' or ``mini-Neptunes''; here we use
the term ``super-Earth.''}
The {\it Kepler} mission
showed that more than half of FGKM dwarfs have super-Earth
companions out to orbital periods of $\sim$100 days \citep[e.g.,][]{fressin13, dressing15charbonneau}.
These planets generally weigh less than 
$20 \me$, as determined from transit-timing analyses 
\citep[e.g.,][]{jontofhutter14, wu13} and radial velocity measurements 
\citep[e.g.,][]{weiss14, marcy14, dumusque14, dressing15}. Their
compositions, as inferred from their bulk densities, span a wide range;
some are consistent with pure rock while others have 1--10\% of their
mass in voluminous H/He envelopes
\citep[e.g.,][]{lopez14, rogers15, lee16chiang}.

Nearly all such planets are detected
orbiting 
Gyr-old host stars.
To better understand how super-Earths form, we need to find
them in younger systems, while they are still emerging
from their parent circumstellar disks.
Unfortunately, the two most prolific planet detection methods,
the radial velocity and transit techniques, are
not suited to detecting super-Earths in disks.
Pre-main-sequence stars tend to be too spectroscopically
and photometrically noisy
because of magnetic activity and
ongoing, time-variable accretion.
An exception is provided by K2-33b, an $\rplanet=6\re$ planet
discovered transiting a 5--10 Myr old star \citep{david16}.
This object may be a super-Earth that will eventually cool
and contract to a radius $\rplanet < 4 \re$.
Radial velocity measurements of its mass are not, however,
strongly constraining ($M_{\rm p} < 3.6 M_{\rm J}$, where
$\mj = 318 \me$ is the mass of Jupiter).
Another example of a young planet, discovered
using radial velocities, is V830 Tau b,
but it is a comparatively rare gas giant, not a super-Earth  
($\mplanet = 0.77 \pm 0.15 \mj$;
\citealt{donati16}). 
As far as direct imaging goes,
gas giants a few dozen Myr old have been
detected with high-contrast
near-infrared (NIR) adaptive optics systems
(Gemini/GPI and VLT/SPHERE):
see 51 Eri b \citep{macintosh15}
and HD 131399Ab \citep{wagner16}.
The current detection limit of such instruments
is $\sim$1$\mj$ for planets a few Myr old
\citep[e.g.,][]{bowler16}.
Directly imaging super-Earths is still out of
reach.

On the other hand, these same NIR
adaptive optics technologies,
in combination with the Atacama Large Millimeter Array (ALMA),
have enabled us to detect structures in disks produced
by gravitational interactions with embedded planets,
and thus to observe, indirectly, planet formation in action.
Scattered light images in the NIR and thermal emission maps
at millimeter (mm) wavelengths
have achieved angular resolutions less than $0\arcsec.05$,
corresponding to length scales of a few AUs at the locations
of nearby star-forming regions such as Taurus.
Disk-planet interactions can produce a variety of structures, depending on
the planet mass $\mplanet$, the disk scale height $h$, and the
disk viscosity as parameterized by $\alpha$, the ratio of the viscous
stress to the thermal pressure \citep{shakura73}.
So far, three types of possible planet-induced structures have been
possibly detected in disks:
\begin{enumerate}
\item Gaps (in, e.g., HD 169142; \citealt{quanz13gap, momose15, fedele17}).
\item Density waves (in, e.g., AB Aur; \citealt{hashimoto11, tang17}).
\item Vortices created by the Rossby wave instability 
(RWI; \citealt{lovelace99, li01, li05}) at the edges of planet-opened gaps
(in, e.g., IRS 48 and HD 142527; \citealt{vandermarel13, casassus13}).
\end{enumerate}
In this paper, we expand upon the first of these disk features,
showing how a single planet can actually carve out two gaps to either
side of itself, separated by a lane of co-orbital material.
This is the signature of a low mass planet in a low viscosity or inviscid
disk. By ``low mass,'' we mean a planet whose mass is less than the thermal mass
\begin{equation}
\mth \simeq M_\star\left(\frac{h}{r}\right)^3 \simeq 40 \me \left( \frac{\mstar}{\msun} \right) \left(\frac{h/r}{0.05}\right)^3 
\label{eq:mthermal}
\end{equation}
where $M_\star$ is the host stellar mass and $r$ is the planet-star separation. In flared disks, $h/r$ increases
with $r$, and by extension so does $\mth$.
Thermal mass planets excite density waves at their Lindblad
resonances that are already non-linear
(characterized by
fractional
surface density perturbations
$\Delta \Sigma / \Sigma \sim 1$)
at launch. Sub-thermal masses---a category to which super-Earths
belong---generate weaker waves. By ``low viscosity,'' we mean 
that the waves damp less by intrinsic disk viscosity and more by
steepening and eventually breaking. \citet{goodman01} showed that in an inviscid disk
having a two-dimensional (2D) adiabatic index $\gamma$,
waves launched to either side of the planet's orbit
travel a radial distance
\begin{equation}
l_{\rm sh}\approx0.8\left(\frac{\gamma+1}{12/5}\frac{\mplanet}{\mth}\right)^{-2/5}h
\label{eq:lsh}
\end{equation}
before shocking and dissipating \citep[see also][]{rafikov02}.
For planets with $\mplanet/\mth \sim 0.1$--$0.5$
(i.e., $\mplanet = 4$--$20 \me$ with $h/r=0.05$),
we have $l_{\rm sh}\sim 1$--$2h$.
Where the waves break, they deposit the angular momentum
they carry to disk gas.
As a result, material is repelled away from the planet to form two gaps, at radial separations
$\pm l_{\rm sh}$ from the planet's orbit.
Numerical simulations confirm this ``double gap'' feature in inviscid
disks (see, e.g., Fig.~2 in \citealt{dong11nonlinear}
or Fig.~8 in \citealt{duffell12}; see also \citealt{li09, muto10,
  dong11linear, zhu13}).

The double gaps are expected to be shallow.
From the numerical fitting formula of Fung et al.~(\citeyear{fung14}; see also \citealt{duffell13,kanagawa15,
  duffell15gap}), a super-Earth of mass $10 M_\oplus$ orbiting
a $1 M_\odot$ star opens a gap having a surface density contrast of
\begin{equation}
\frac{\Sigma_{\rm gap}}{\Sigma_0} \simeq
0.4 \left(\frac{3\times 10^{-5}}{\mplanet/\mstar}\right)^{2.16} \left(\frac{h/r}{0.05}\right)^{6.61}
\left( \frac{\alpha}{10^{-4}} \right)^{1.41} 
\label{eq:gap}
\end{equation}
where $\Sigma_{\rm gap}$ is the surface density inside the gap
and $\Sigma_0$ is the unperturbed value.
The above formula is known to break down as $\alpha \rightarrow 0$
because hydrodynamic instabilities substitute for viscous diffusion
in filling up disk gaps;
simulations by \citet{fung17}
show that super-Earths open gaps with $\Sigma_{\rm gap}/\Sigma_0 \sim 0.1$
when $\alpha = 0$ and $h/r = 0.03$.

Although gas surface density contrasts generated by super-Earths may be modest,
aerodynamic interactions between gas and dust can produce
higher contrast features in dust
\citep[e.g.,][]{rice06, pinilla12-dusttrapping}.
In particular, the edges of planet-opened gaps 
are regions of high gas pressure,
and dust particles drift toward such regions and collect into overdense rings
\citep[e.g.,][]{paardekooper04, paardekooper06, zhu12,
  pinilla12-diffcavsize}.
Dust rings have been detected \citep[e.g.,][]{zhang14, hashimoto15, vandermarel15, dong17j1604}. 

In this paper, by combining global 2D two-fluid (gas+dust) hydrodynamical
simulations with a radiative transfer code, we study how
super-Earths produce characteristic double-gap patterns
in mm-sized dust particles in nearly inviscid ($\alpha \lesssim
10^{-4}$) disks. Such structures should be readily
detectable in ALMA dust continuum observations of disks 140 pc away
with $\sim$0$\arcsec.03$ angular resolution. We will introduce our
simulations in Section~\ref{sec:simulations}, present results
in Section~\ref{sec:results}, and summarize and discuss in
Section~\ref{sec:summary}.


\section{Simulations}\label{sec:simulations}

We calculate the dust and gas distributions in a disk perturbed by a
planet using global 2D (radial-azimuthal) hydrodynamics simulations
with the {\tt LA-COMPASS} code (\citealt{li05, li09, fu14};
Section~\ref{sec:hydro}). Hydro models are post-processed
using the \citet{whitney13} Monte Carlo radiative transfer (MCRT) code
to produce synthetic images at various wavelengths
(Section~\ref{sec:mcrt}).

\subsection{Hydrodynamics Simulations}\label{sec:hydro}

The simulation setup is largely adopted from \citet{fu14} and
\citet{miranda16}, and is briefly summarized here.
Dust is treated as a pressureless fluid \citep[e.g.,][]{takeuchi02}. An $\alpha$-viscosity is added to the gas dynamics, which introduces an additional diffusion term in the dust continuity equation \citep[see Eqn.~28 in][]{takeuchi02}. We do not model turbulent mixing. 
An axisymmetric 2D global disk is initialized in cylindrical
coordinates ($r\times\phi=1024\times1024$) with a gas surface density
\begin{equation}
\Sigma_{\rm gas,0}=5.56\left(\frac{\rp}{r}\right)^\beta\ {\rm g\ cm^{-2}},
\label{eq:sigmag0}
\end{equation}
where $\beta$ is the global radial power law index ranging from 0 to
1.5, and $\rp=30$ AU is the orbital radius of the planet. The planet's
orbit is coplanar with the disk, circular, and fixed in time
for our default simulations; in Section \ref{sec:migration},
we experiment with prescriptions for radial migration.
The planet mass ramps up with time $t$ as $\mplanet(t)\propto \sin{(t)}$ over $0\leq t \leq \tau_{\rm growth}$. 
Our default setting is $\tau_{\rm growth} = 10$ orbits; in Section~\ref{sec:10me-growth}, we experiment with $\tau_{\rm growth} = 1000$ and 3000 orbits.
The size of the simulation domain is 2$\pi$ in
$\phi$ and 0.1--2.1$\rp$ in $r$ (3--64 AU). At the end of the hydro
simulations, before MCRT post-processing, the inner disk
within twice the inner boundary is manually removed to avoid
possible boundary condition artifacts. Most models take $\beta=1$, for which
the initial gas mass is about $0.01\msun$ within 60
AU. The gas surface density and velocity are fixed at their initial
values at both the inner and outer boundaries. 

The equation of state of the gas  
is locally isothermal. The scale height of the
gas disk $h$ is taken as
\begin{equation}
\frac{h}{r}=0.05\left(\frac{r}{\rp}\right)^{0.25} \,;
\label{eq:h/r}
\end{equation}
the disk is flared. The central star is $1\msun$ in mass, and the
thermal mass at $\rp$ is $0.05^3\times\msun=40\me$
(Eqn.~\ref{eq:mthermal}). The gas has a kinematic viscosity
$\nu=\alpha h^2\Omega_{\rm k}$ \citep{shakura73} with constant
$\alpha$ and $\Omega_{\rm k}$ equal to the Keplerian angular velocity.
We experiment with $\alpha$ between $5\times10^{-4}$ and
$5\times10^{-6}$.

Dust particles with size $s$ are treated as a second fluid in
the hydro simulation. The Stokes number St (momentum stopping
time normalized to the dynamical time) of dust particles in
the Epstein regime is
\begin{equation}
{\rm St}=\frac{\pi s \rho_{\rm dust}}{2\sigmag},
\label{eq:st}
\end{equation}
where $\rho_{\rm dust}$ is the internal density of a dust particle. At
$t=0$ and $r=\rp$, ${\rm St}=0.007$ for $\rho_{\rm dust}= 1.2$ g cm$^{-3}$
and $s=0.2$ mm (``big dust''; see Section~\ref{sec:mcrt}).
The two-fluid hydrodynamic equations describing the coupled
evolution of gas and dust are solved in LA-COMPASS including both aerodynamic drag
on dust, and dust back-reaction on gas.
The big dust particles are initialized with a surface density
linearly proportional to the gas,
$\Sigma_{\rm big-dust,0}=0.5\%\times\Sigma_{\rm gas,0}$.
An outflow inner boundary condition and an inflow outer boundary
condition are imposed on the dust. 

We summarize the models and their parameters in
Table~\ref{tab:models}. Our default is to run for 1500 orbits (0.25 Myr
at 30 AU); in Section~\ref{sec:evolution}, we experiment with longer run
times.

\subsection{Radiative Transfer Simulations and Synthetic Observations}\label{sec:mcrt}

We follow the procedures described in \citet{dong15gap} to translate
the hydro disk models to synthetic observations. The main steps
are summarized here.
Light from the central star is scattered and
absorbed 
by ``small dust'' particles at the disk surface.
The small dust is assumed to
resemble sub-$\mu$m-sized, interstellar medium (ISM) 
dust in its optical properties (\citealt{kim94}; see Fig. 2 in \citealt{dong12cavity}).
Because of their short stopping times,
small dust grains are assumed to be always well mixed with gas
at a mass fraction of 0.5\%.
Big dust particles, having sizes $s = 0.2$ mm
as in the hydro simulations,
are assumed to have settled to the disk midplane
\citep[e.g.,][]{dullemond04dustsettling, birnstiel10}.
Initially in the hydro simulations, the big-to-small-dust mass ratio is 1:1,
so that the initial total-dust-to-gas mass ratio is 1:100;
at the end of the hydro simulation, the big-dust-to-gas ratio in a given
gas parcel has evolved because of drag forces.
The optical properties of the big dust are calculated using the \citet{bohren83} Mie routine,\footnote{The B.~T.~Draine version, https://www.astro.princeton.edu/$\sim$draine/scattering.html, wrapped into a Python package by
T. Robitaille, https://github.com/hyperion-rt/bhmie.}
assuming an ISM composition.
The Whitney MCRT code calculates the equilibrium temperature of all dust
particles using iterative methods \citep{lucy99}.

We assume the central source
is a luminous blackbody sphere of radius
$R_\star=2.3\rsun$ and $T=4350$K,
appropriate for a 1 $\msun$ star at 1 Myr \citep{baraffe98}.
We manually fill the inner disk by using the initial surface density
profile (Eqn.~\ref{eq:sigmag0}) to extrapolate 
surface densities
inward from $r=0.2\rp$.
The hydro simulations are 2D and yield only the surface densities
$\sigmasd$ and $\sigmabd$; these are made into 3D density fields
for use in the MCRT calculations
by puffing them up assuming Gaussian profiles in the vertical direction $z$:
\begin{align}
\rho_{\rm small-dust}(z)&=\frac{\sigmasd}{\hsd\sqrt{2\pi}} e^{-z^2/2\hsd^2}, \label{eq:rhosd}
\\ \rho_{\rm big-dust}(z)&=\frac{\sigmabd}{\hbd\sqrt{2\pi}} e^{-z^2/2\hbd^2}, \label{eq:rhORd}
\end{align}
where $\hsd$ is set equal to the gas scale height $h$ (Eqn.~\ref{eq:h/r}),
and $\hbd = 0.1\hg$ to mimic dust settling, as in \citet{miranda16}.

We produce both $H$-band polarized intensity scattered light images and ALMA band
7 (870 $\micron$; 345 GHz; 7.5 GHz bandwidth) continuum observations,
assuming the disk is 140 pc away at Dec=+25$^\circ$ (e.g, Taurus).
Unless otherwise noted, we assume 4 hours of integration time for simulated ALMA observations.
For $H$-band images, we take an inner working angle of $0\arcsec.07$ (10 AU)
and convolve the raw MCRT products with a Gaussian
point spread function (PSF) having a
full-width-half-maximum
of $0\arcsec.04$ (6 AU) to simulate
the performances of VLT/SPHERE and Gemini/GPI.
For mm dust continuum observations, we transform the raw MCRT products to simulated
ALMA observations using the {\tt simobserve} and {\tt simanalyze} tools in
Common Astronomy Software Applications (CASA). A full array of 50 12-m antennas
is used.
We mostly use the array configuration {\tt alma.out24.cfg},\footnote{\url{https://casaguides.nrao.edu/index.php?title=Antenna_Configurations_Models_in_CASA
}.} achieving a beam size of $0\arcsec.029\times0\arcsec.020$ with the major axis of the beam in the north-south direction
(by comparison, the smallest beam achievable for a Taurus target at 345 GHz with the full ALMA array is $0\arcsec.017\times0\arcsec.012$ with configuration {\tt alma.out28}). Simulated ALMA observations are corrupted by thermal noise assuming the default atmospheric model with 0.9 mm precipitable water vapor ($3^{\rm rd}$ octile). Four hours of integration reaches a root-mean-squared (RMS)
sensitivity of 13 $\mu$Jy beam$^{-1}$ using the ALMA sensitivity calculator.\footnote{\url{https://almascience.eso.org/proposing/sensitivity-calculator}} Throughout the paper, we use a gray color scheme for surface density, a blue-hot color scheme in the $H$-band, and a red-hot color scheme for ALMA images. 
All scattered light images are for face-on disks, while for ALMA images we experiment with non-zero disk inclinations.


\section{Results}\label{sec:results}

After presenting a fiducial model that illustrates
the basic observable disk features produced by a super-Earth
(Section~\ref{sec:10me}), we
then explore how these features vary with planet mass (Section~\ref{sec:mp}),
disk viscosity (Section~\ref{sec:viscosity}),
background surface density profile (Section~\ref{sec:sigma0}),
and orbital migration of the planet (Section~\ref{sec:migration}).

\subsection{A 10-$\me$ Super-Earth at 30 AU}\label{sec:10me}

We use model 10ME (see Table 1)
to showcase the basic disk features produced by a super-Earth
in a low viscosity environment.

\subsubsection{Surface Density Distributions}\label{sec:10me-density}

\begin{figure}
\begin{center}
\includegraphics[trim=0 0 0 0, clip,width=0.6\textwidth,angle=0]{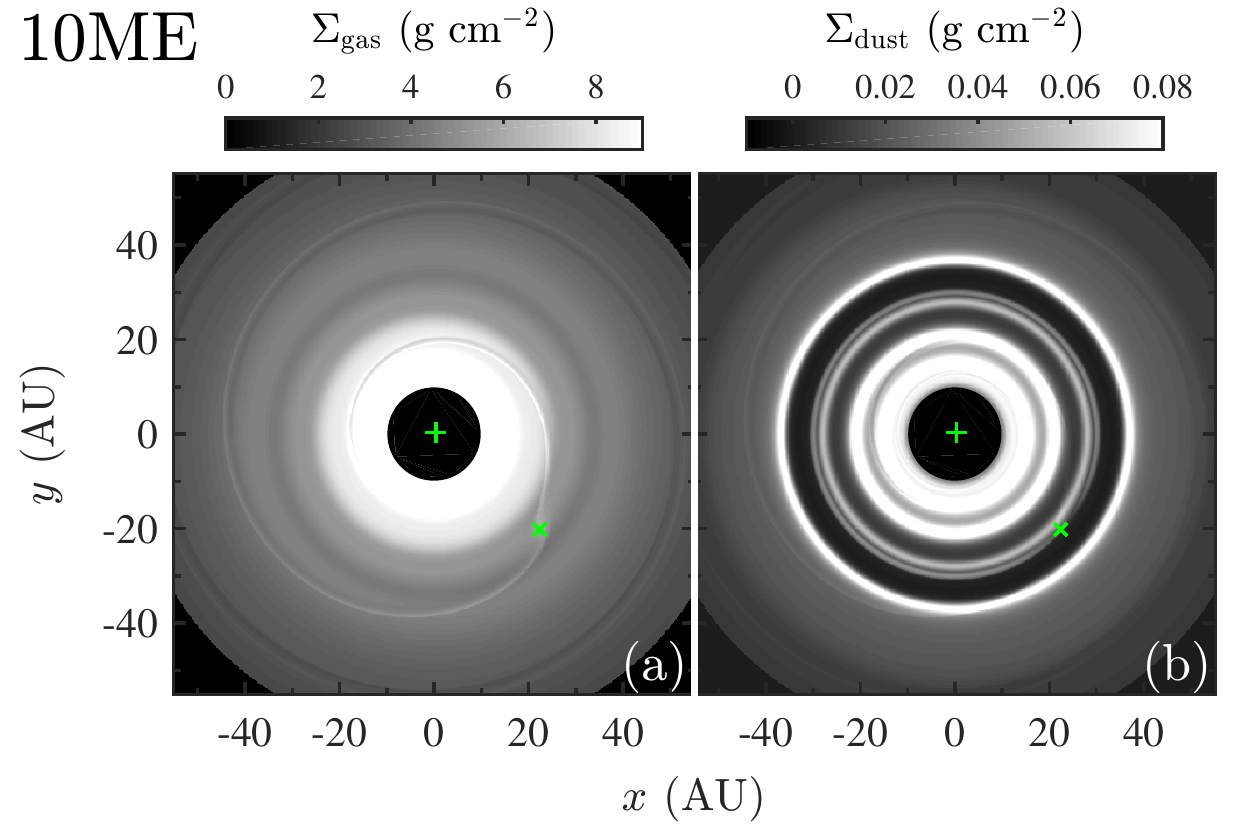}
\includegraphics[trim=0 0 0 0, clip,width=0.5\textwidth,angle=0]{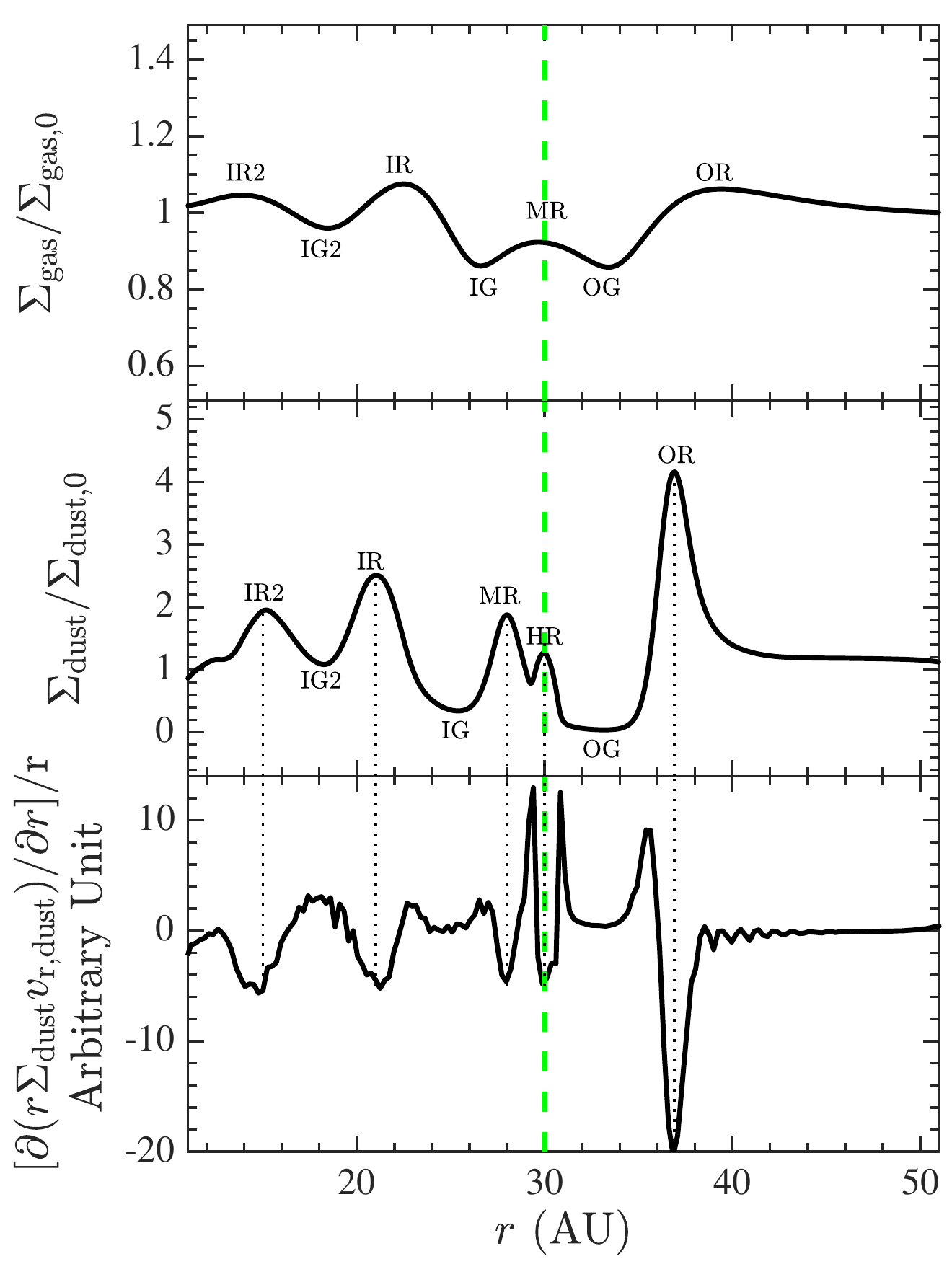}
\end{center}
\figcaption{Surface density maps ({\bf top}) and their azimuthally-averaged radial profiles ({\bf bottom}) of our fiducial model 10ME (10 $\me$, $\alpha=5\times10^{-5}$) at 1500 orbits (0.25 Myr). The inner 8 AU is masked out. The green ``$+$'' and ``$\times$'' symbols in (a) and (b) mark the locations of the star and the planet, respectively. The vertical dashed line running through the radial profile panels indicates the orbital radius of the planet ($\rp=30$ AU). The major features on the two radial profiles are labeled: IR2 (inner ring 2), IG2 (inner gap 2), IR (inner ring), IG (inner gap), MR (middle ring), OG (outer gap), and OR (outer ring). There is in addition the horseshoe ring (HR) seen in dust only. One super-Earth can produce multiple rings and gaps in the distributions of gas and of $\sim$mm-sized dust. See Section~\ref{sec:10me} for details.
\label{fig:sigma_10me}}
\end{figure}

Figure~\ref{fig:sigma_10me} shows the surface density maps of the gas and of the $0.2$-mm-sized big dust, as well as their azimuthally averaged radial profiles. Henceforth, for convenience, we sometimes omit ``big'' when referring to ``big dust.'' In the gas, the 10-$M_\oplus$ super-Earth gradually opens two narrow gaps about $2h$ away from its orbit, as predicted by \citet{rafikov02migration} in this nearly inviscid environment. The gas in the coorbital region is less depleted than in the two gaps, and forms an annular ring. We identify six features in the gas radial profile: apart from the IG (inner gap), OG (outer gap), and MR (middle ring), we observe an IR (inner ring), an OR (outer ring), an IR2 (inner ring 2), and an IG2 (inner gap 2). The gas originally in the gap regions is pushed away and piled up at the gap edges to form the IR and OR. Because of the low viscosity, viscous diffusion is unable to smooth away the two rings at the time shown ($t = 1500$ orbits). These structures in the gas surface density have been seen in previous hydro simulations of low mass planets in low viscosity disks (see, e.g., Figure 2 of \citealt{li09}, Figure 14 of \citealt{muto10}, and Figure 8 of \citealt{duffell12}). The IR2 feature (and an accompanying annular gas depletion IG2) has been seen in simulations with inner boundaries smaller than $\sim$0.4$\rp$ 
(see, e.g., Figure 2 of \citealt{li09}, Figure 9 of \citealt{yu10}, and Figure 1 of \citealt{zhu13}).

Dust drifts relative to gas. In a 2D disk, ignoring azimuthal gradients 
and a small diffusion term, the continuity equation for the dust reads
\begin{equation}
\frac{\partial \sigmad}{\partial t}+\frac{1}{r}\frac{\partial(r\sigmad v_{r,\rm dust})}{\partial r}\approx0,
\label{eq:cont}
\end{equation}
where 
\begin{equation}
v_{r,\rm dust}=v_{r,\rm gas}+\frac{t_{\rm s}}{\sigmag}\frac{\partial p}{\partial r}
\label{eq:vrdust}
\end{equation}
is the radial drift velocity of dust,
$p$ is gas pressure, and $t_{\rm s}= {\rm St}/\Omega_{\rm k}$
is the dust stopping time.
Dust will pile up at locations where the second term in (\ref{eq:cont}) is most negative. This term is plotted at the bottom of Figure~\ref{fig:sigma_10me}. In total, 5 local minima can be identified, each corresponding to a dust concentration. The big-dust-to-gas mass ratio, initially equal to 0.5\% everywhere, increases at the locations of dust rings by factors of up to 4.  The horseshoe dust ring HR (seen before in, e.g., Figure 4 of \citealt{zhu14votices}) has no obvious counterpart in $\sigmag$ and is caused by a different mechanism --- 1:1 gravitational resonant interaction with the planet.

The dust rings are not necessarily located at the positions of low-order mean motion resonances relative to each other, although they may appear to be close to resonances by chance. The locations of the rings
can change with planet mass, background disk profile, or planet
migration, as will be seen later.

The behavior of dust in a disk with a sub-Jupiter mass planet has been investigated by many others \citep[e.g.,][]{paardekooper04,paardekooper06,lyra09, zhu14votices, rosotti16} using two-fluid simulations. Our results generally agree with these works in the low $\mplanet$ and low $\alpha$ limit (see, e.g., the multiple gaps and rings in the 0.05$\mj$ model in Figure 8 of \citealt{paardekooper06}, and the 8$\me$ model in Figure 4 of \citealt{zhu14votices}). Note that our 10$\me$ planet is not massive enough to trap dust effectively at its 2:1 resonance at $r=1.6 r_{\rm p}$, by contrast to the higher mass planet considered by \citet{paardekooper04, paardekooper06}.

Our planet excites density waves both inside and outside its orbit. The pressure structures in the density waves are not able to concentrate dust particles via dust-gas coupling because the waves corotate with the planet and not with the local disk. Dust particles only spend a small fraction of time inside the pressure overdensities associated with waves. By contrast, density waves excited by gravitational instability do rotate at approximately the local Keplerian velocity and are able to concentrate particles of certain sizes \citep{rice04}.

\subsubsection{Synthetic Observations and Detectability}\label{sec:10me-image}

\begin{figure}
\begin{center}
\includegraphics[trim=0 0 0 0, clip,width=0.6\textwidth,angle=0]{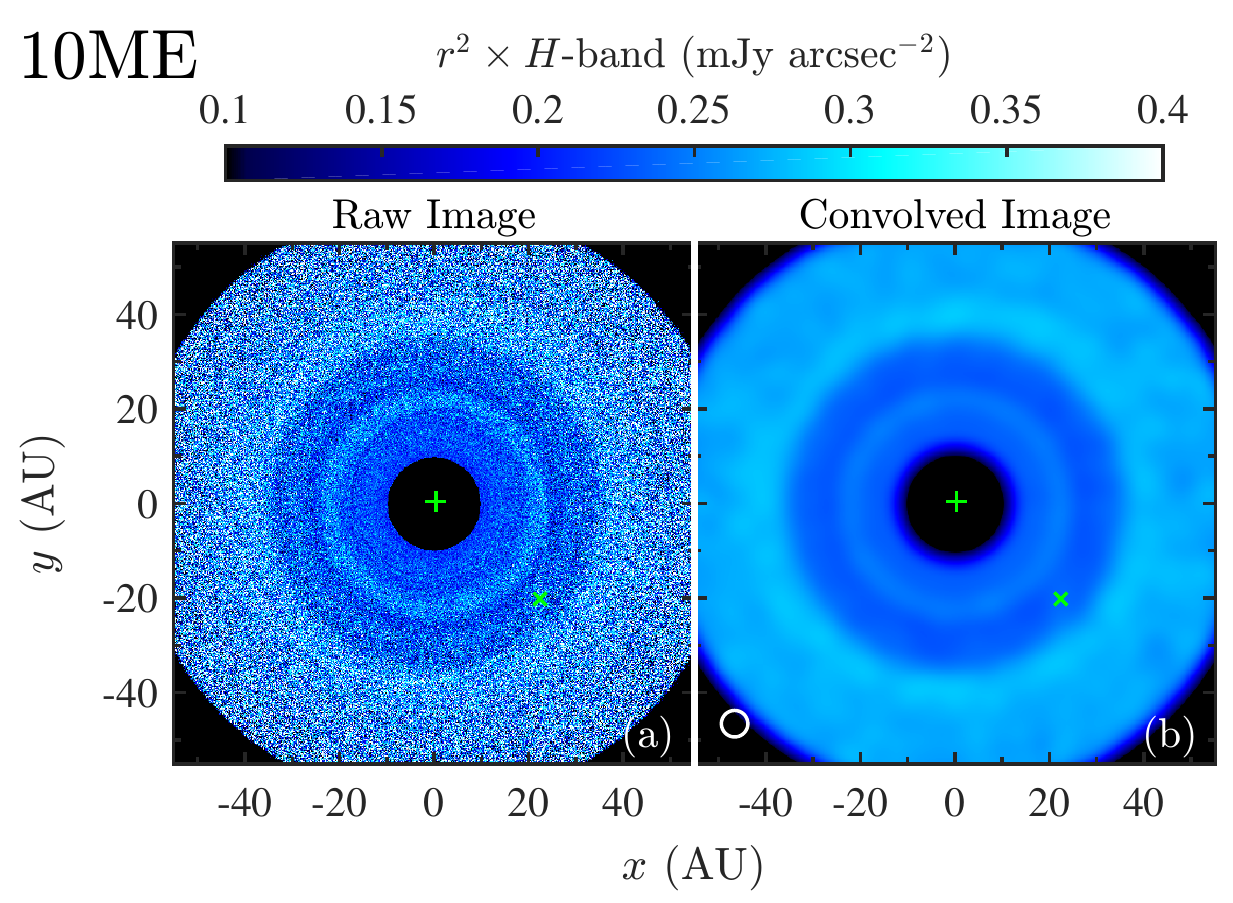}
\includegraphics[trim=0 0 0 0, clip,width=0.39\textwidth,angle=0]{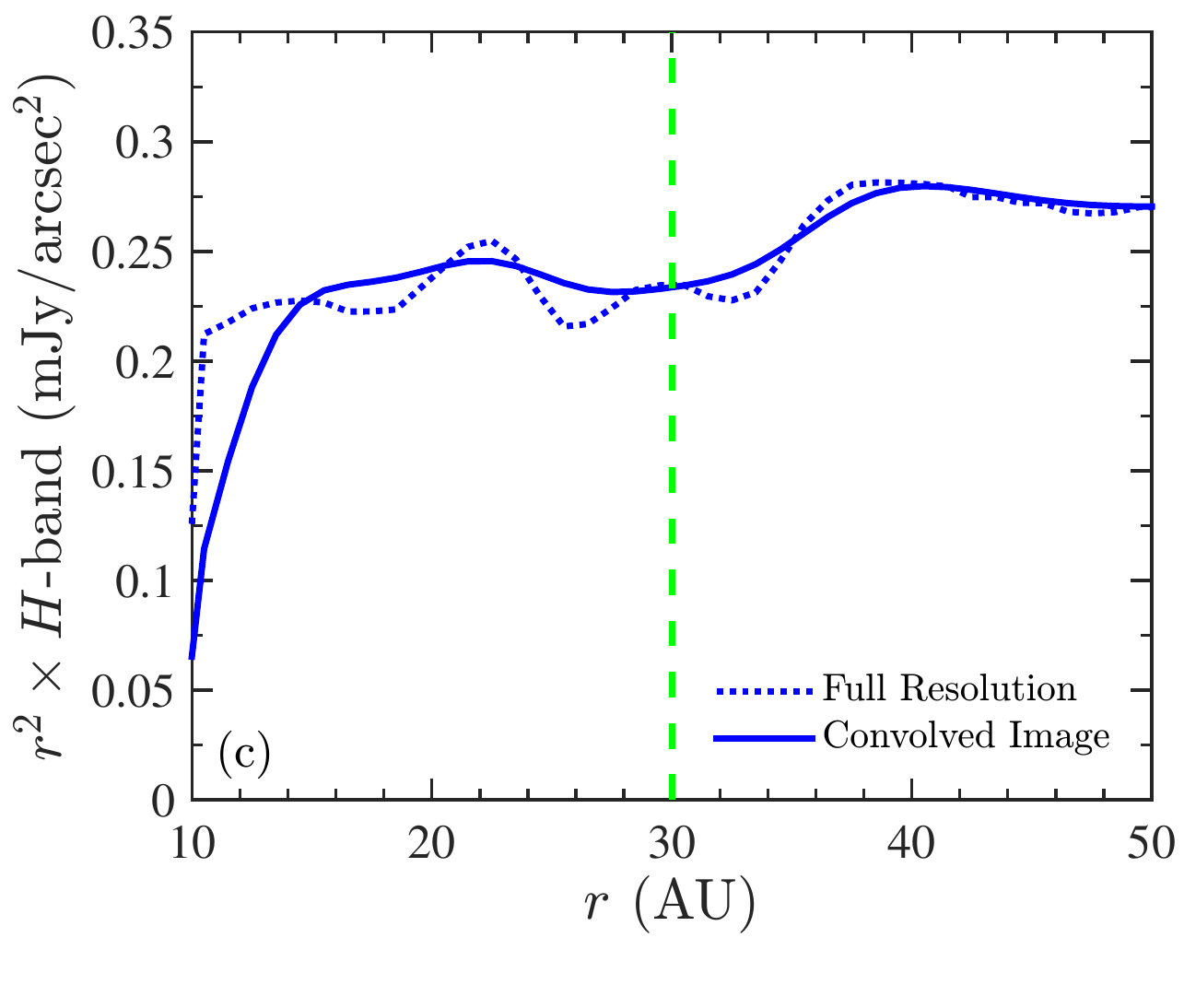}
\includegraphics[trim=0 0 0 0, clip,width=0.6\textwidth,angle=0]{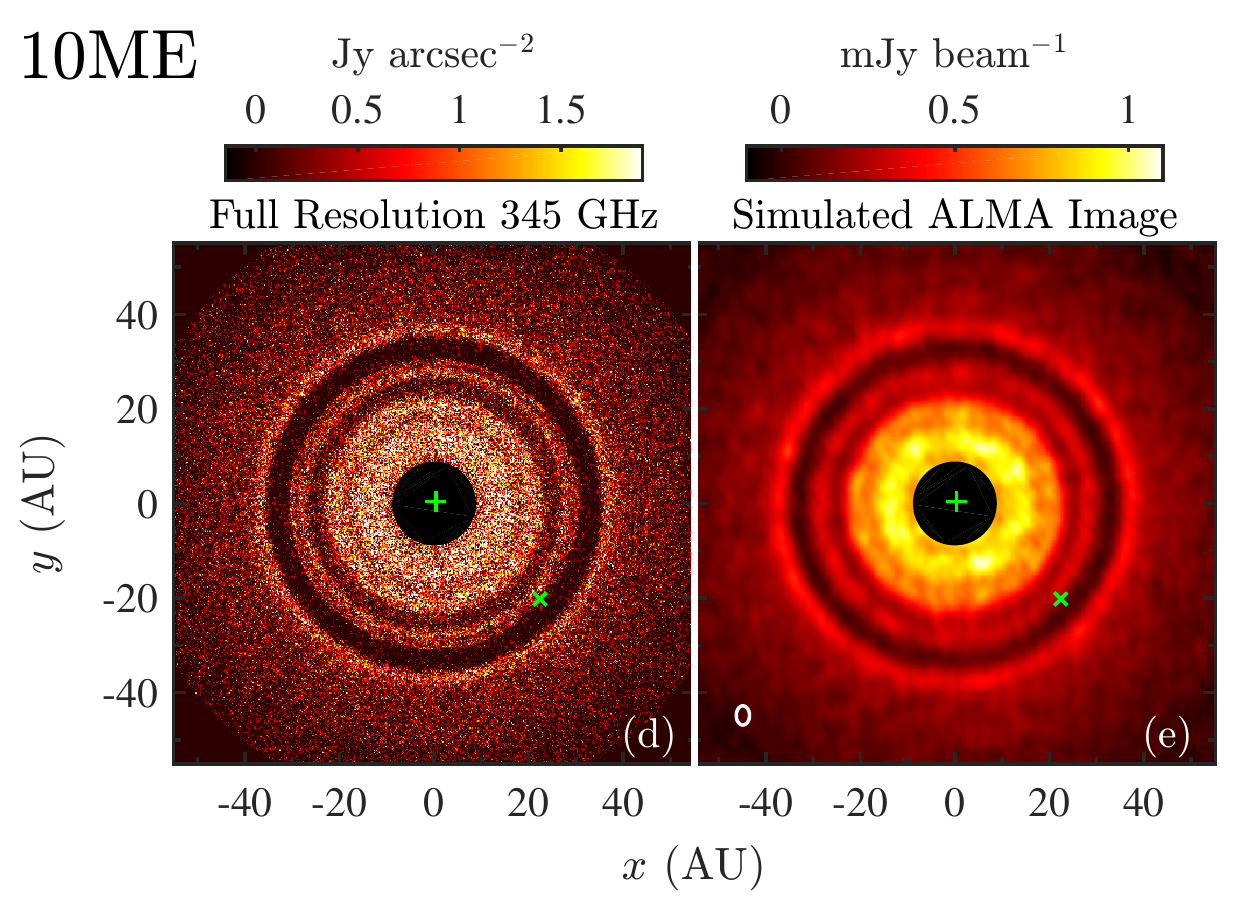}
\includegraphics[trim=0 0 0 0, clip,width=0.39\textwidth,angle=0]{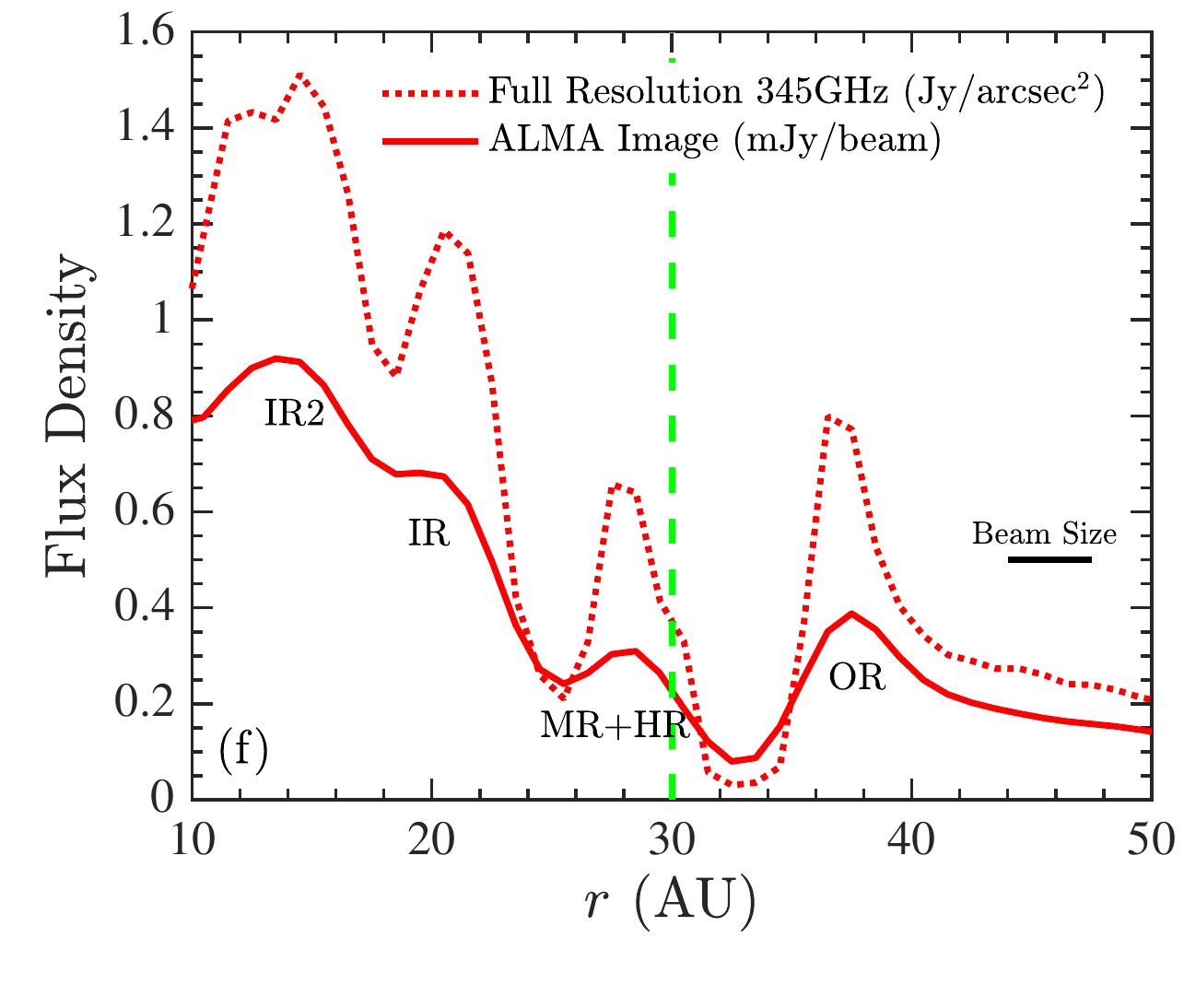}
\end{center}
\figcaption{$H$-band images ({\bf top row}) and 0.87 mm continuum images ({\bf bottom row}), and their azimuthally averaged radial profiles, for model 10ME (10 $\me$, $\alpha=5\times10^{-5}$) at 1500 orbits (0.25 Myr). {\bf Panels (a) and (d)} are the full resolution radiative transfer images. {\bf Panel (b)} is panel (a) convolved by a Gaussian PSF (lower left corner) to achieve an angular resolution of $0\arcsec.04$. {\bf Panel (e)} is the simulated ALMA Band 7 image produced from panel~(d) with a beam size of $0\arcsec.03\times0\arcsec.02$ (lower left corner). The inner $0\arcsec.07$ (10 AU) in the NIR images is masked out to mimic an inner working angle, and the inner 8 AU in ALMA images is masked out to avoid inner boundary artifacts in the hydro simulations. The $H$-band images have been $r^2$-scaled. The vertical dashed lines in {\bf panels (c) and (f)} mark the orbital radius of the planet. Four rings can be discerned in the mm continuum images, and are labeled in (f). See Section~\ref{sec:10me} for details.
\label{fig:image_10me}}
\end{figure}

Figure~\ref{fig:image_10me} shows simulated $H$-band polarized intensity and ALMA Band 7 images. At $H$-band, the depletion of the gas (and thus of the entrained small dust) in the two gaps results in two shallow gaps in the full resolution image. After PSF convolution, however, the two gaps merge to form an apparent single gap. Depending on the noise level and the absolute brightness of the disk, such a shallow gap may or may not be detectable in a real system. In the full resolution mm dust
continuum image, four dust rings can be distinguished: IR2, IR, OR, and a merged MR+HR. The simulated ALMA observation recovers nearly 100\% of the total flux, and generally preserves the main features of the $\sigmad$ radial profile. In particular, in both the image and the radial profile, the signature of a super-Earth in a low viscosity disk --- a pair of gaps (a ``double gap'') separating three rings --- is distinctive. 

\begin{figure}
\begin{center}
\includegraphics[trim=0 0 0 0, clip,width=0.9\textwidth,angle=0]{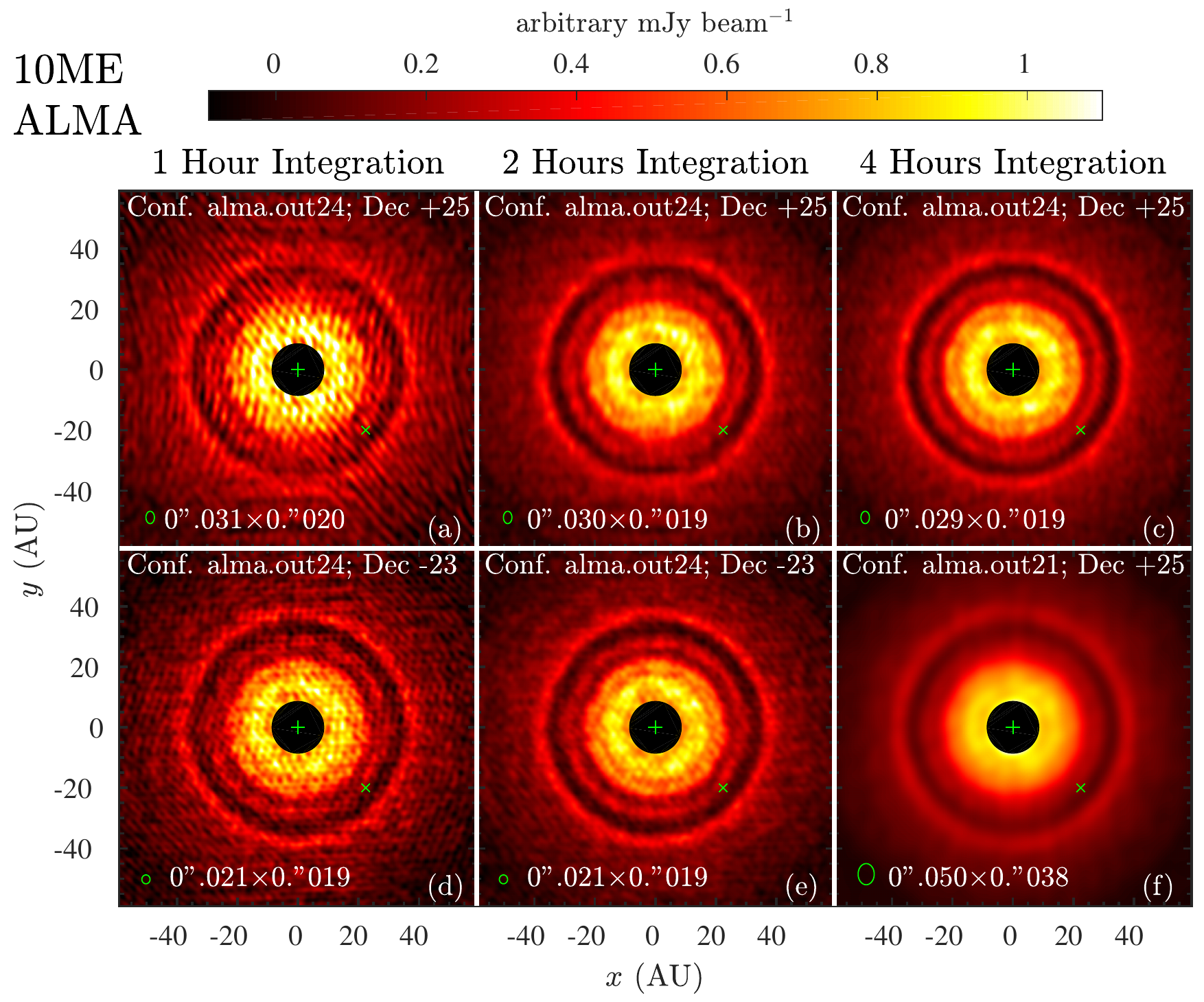}
\end{center}
\figcaption{Dust rings and gaps in model 10ME (10 $\me$, $\alpha=5\times10^{-5}$) at 1500 orbits (0.25 Myr) as observed by ALMA under different conditions. The inner 8 AU in all panels is masked out to avoid inner boundary artifacts in the hydro simulations. {\bf Panels (a)--(c)} show the effects of increasing the integration time from 1 hour to 4 hours (panel (c) is same as panel (e) in Figure~\ref{fig:image_10me}), assuming the target is in Taurus (Dec. $+25^\circ$). {\bf Panel (d) and (e)} assume 1 and 2 hours integration, respectively, and a southern target (Dec. $-23^\circ$). {\bf Panels (a)--(e)} use array configuration {\tt alma.out24.cfg} (beam size indicated at the bottom). {\bf Panel (f)} assumes the same observing conditions as panel (c) but uses array configuration {\tt alma.out21.cfg} (which is more
compact
and therefore has a larger beam size). See Section~\ref{sec:10me-image} for details.
\label{fig:almaimage_10me_detectability}}
\end{figure}

Figure~\ref{fig:almaimage_10me_detectability} illustrates the detectability of gaps and rings under various observational conditions using ALMA. Assuming the disk is at Taurus (dec. $\sim25^\circ$) and using array configuration {\tt alma.out24.cfg} (angular resolution $0\arcsec.03\times0\arcsec.02$), the distinctive double-gap feature can be clearly resolved in 2 hours of integration time (panel (b); RMS noise level 18 $\mu$Jy beam$^{-1}$; S/N$\sim$22 at the OR), and is marginally detectable with 1 hour of integration time (panel (a)). With 4 hours of integration (panel (c)) all 4 rings can be detected with S/N$>$20 (although the separation of IR2 and IR is marginal).
If the object is in the southern sky (dec. $-23^\circ$, panels (d)--(e)), the
double-gap feature is more easily resolved because of better {\it uv}
coverage. 
For array configuration {\tt alma.out21.cfg} which yields a coarser
angular resolution of $0\arcsec.05\times0\arcsec.04$, the double-gap
structure becomes difficult to discern (panel (f)).

\begin{figure}
\begin{center}
\includegraphics[trim=0 0 0 0, clip,width=0.9\textwidth,angle=0]{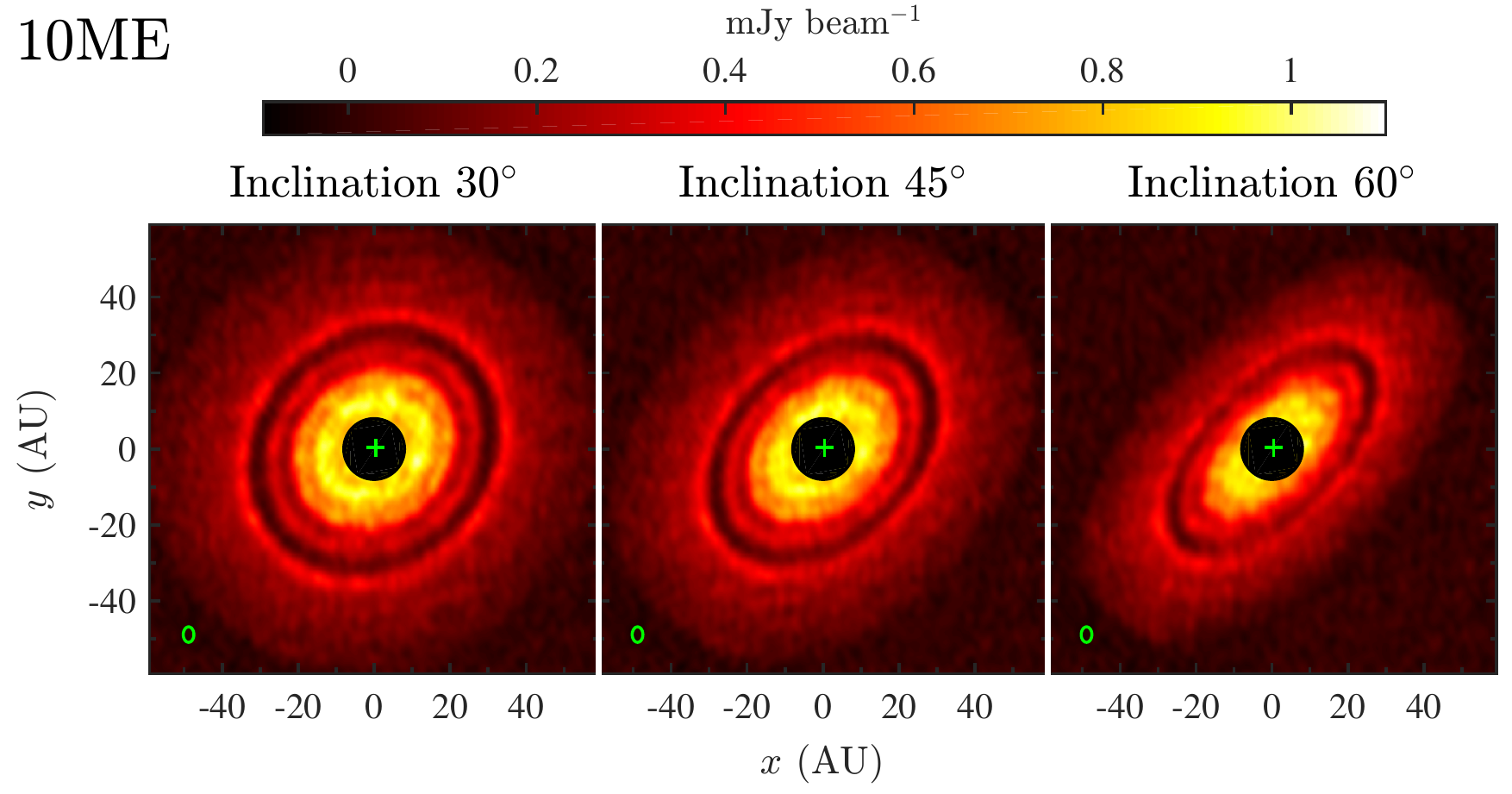}
\end{center}
\figcaption{Similar to Figure~\ref{fig:almaimage_10me_detectability}, but for different disk inclinations. In all panels the disk is at a position angle of $45^\circ$. The double-gap feature can be marginally recognized at an inclination of $60^\circ$ (along the major axis). See Section~\ref{sec:10me-image} for details.
\label{fig:almaimage_10me_inclination}}
\end{figure}

Figure~\ref{fig:almaimage_10me_inclination} shows the morphology of the
rings and gaps in ALMA observations at various disk inclinations. In
all cases shown, the disk has a position angle of $45^\circ$, in between
the directions of the major and minor axes of the beam. At $45^\circ$
disk inclination, the structures in the disk can still be detected, particularly the double gap; at $60^\circ$ inclination, structures along the minor axis of the disk threaten to be lost, while
rings and gaps can still be seen along the major axis.

Our hydro simulations are 2D and are thus not suitable for simulating the
appearance of density waves in scattered light, as these waves have
substantive vertical motions \citep{zhu15densitywaves}. However, as
judged
from the 3D hydro and radiative transfer simulations of
\citet{dong17spiralarm}, the density waves excited by low mass planets 
seem generally too weak to be detected
by current NIR imaging facilities. 

\subsubsection{Time Dependence}\label{sec:evolution}

\begin{figure}
\begin{center}
\includegraphics[trim=0 0 0 0, clip,width=\textwidth,angle=0]{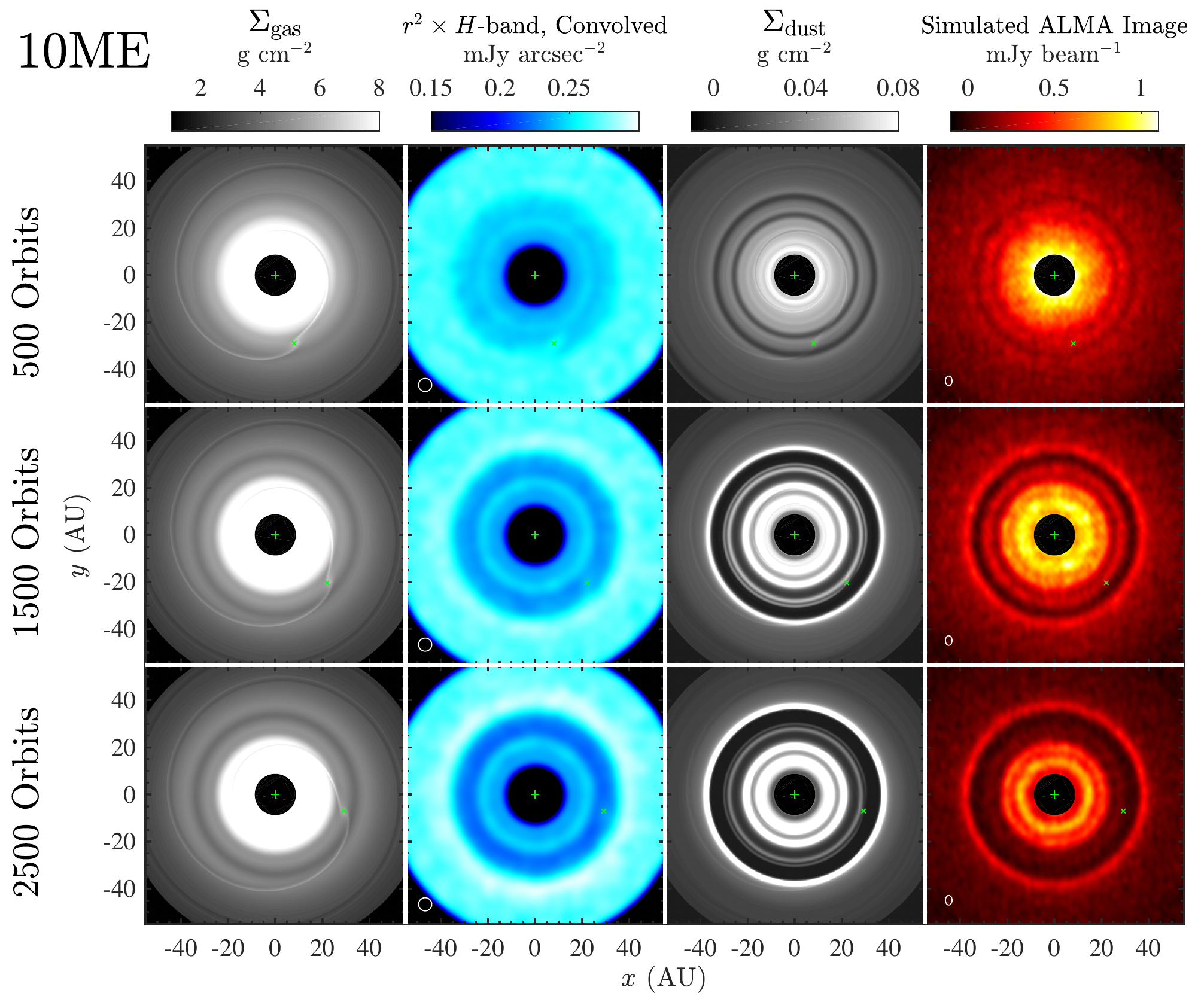}
\end{center}
\figcaption{Surface densities of small ({\bf left column}) and big ({\bf $3^{\rm rd}$ column}) dust particles; the convolved $H$-band image ({\bf $2^{\rm nd}$ column}); and the simulated ALMA Band 7 image ({\bf right column}) of model 10ME at 500 orbits (80k years; {\bf top}), 1500 orbits (250k years; {\bf middle}), and 2500 orbits (410k years; {\bf bottom}). The double-gap feature takes time to develop from our assumed smooth initial conditions and weakens at late times, but can be detected over timescales of thousands of orbits. See Section~\ref{sec:evolution} for details.
\label{fig:image_times}}
\end{figure}

Figure~\ref{fig:image_times} shows the time evolution of the disk
(azimuthally averaged radial profiles are shown in the last column of
Figure~\ref{fig:rp}). 
The observational signatures noted above are transient; in particular,
the central ring (MR+HR) seen in the mm continuum gradually diffuses,
merging the two gaps into one and eventually
killing the signature of a double gap. Nevertheless, the double-gap
feature can still last long enough to be detectable,
about 2000 orbits (0.3 Myr for $r_{\rm p} = 30$ AU) for the model
shown.

\subsubsection{Lengthening the Planet Growth Timescale}\label{sec:10me-growth}

\begin{figure}
\begin{center}
\includegraphics[trim=0 0 0 0, clip,width=0.49\textwidth,angle=0]{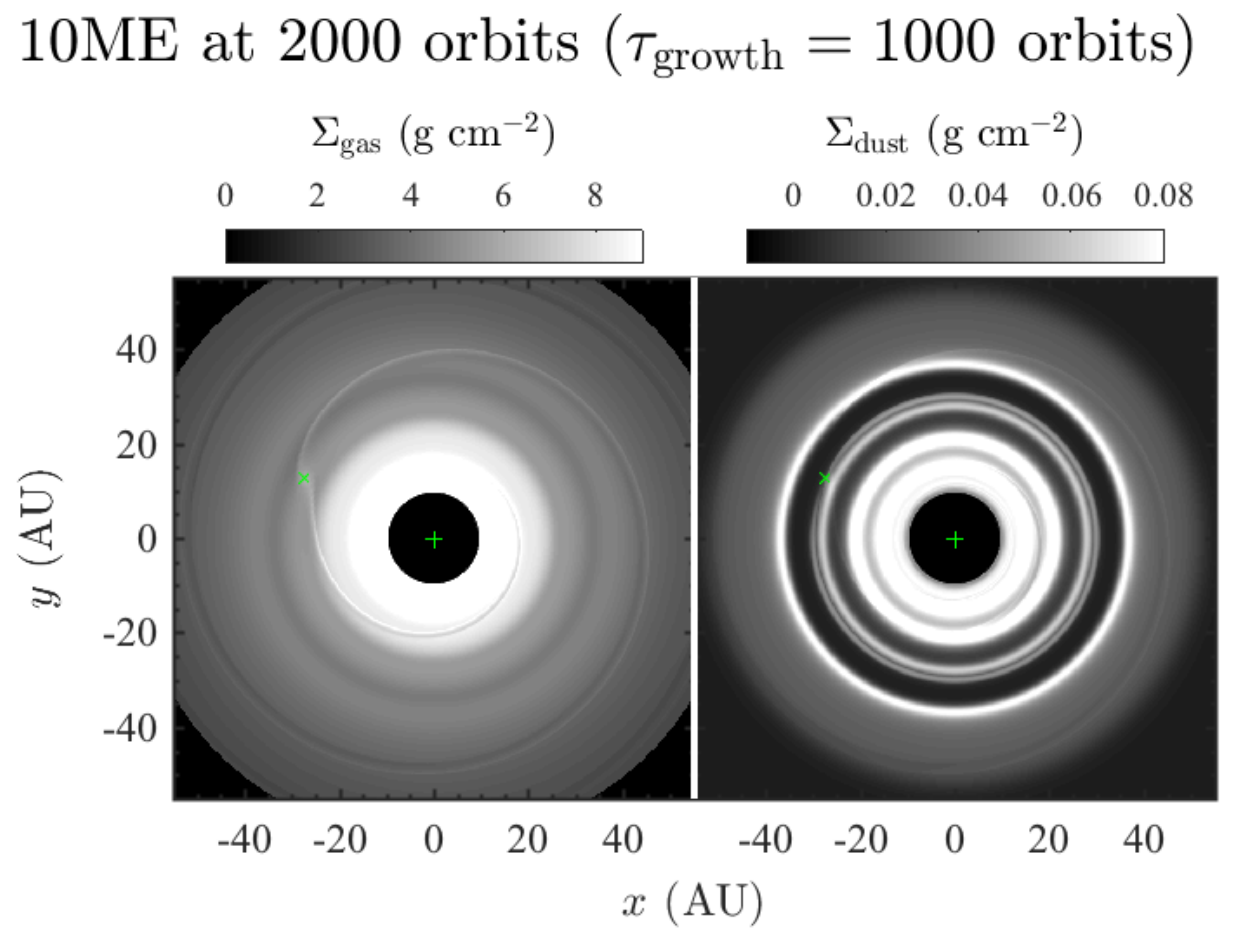}
\includegraphics[trim=0 0 0 0, clip,width=0.49\textwidth,angle=0]{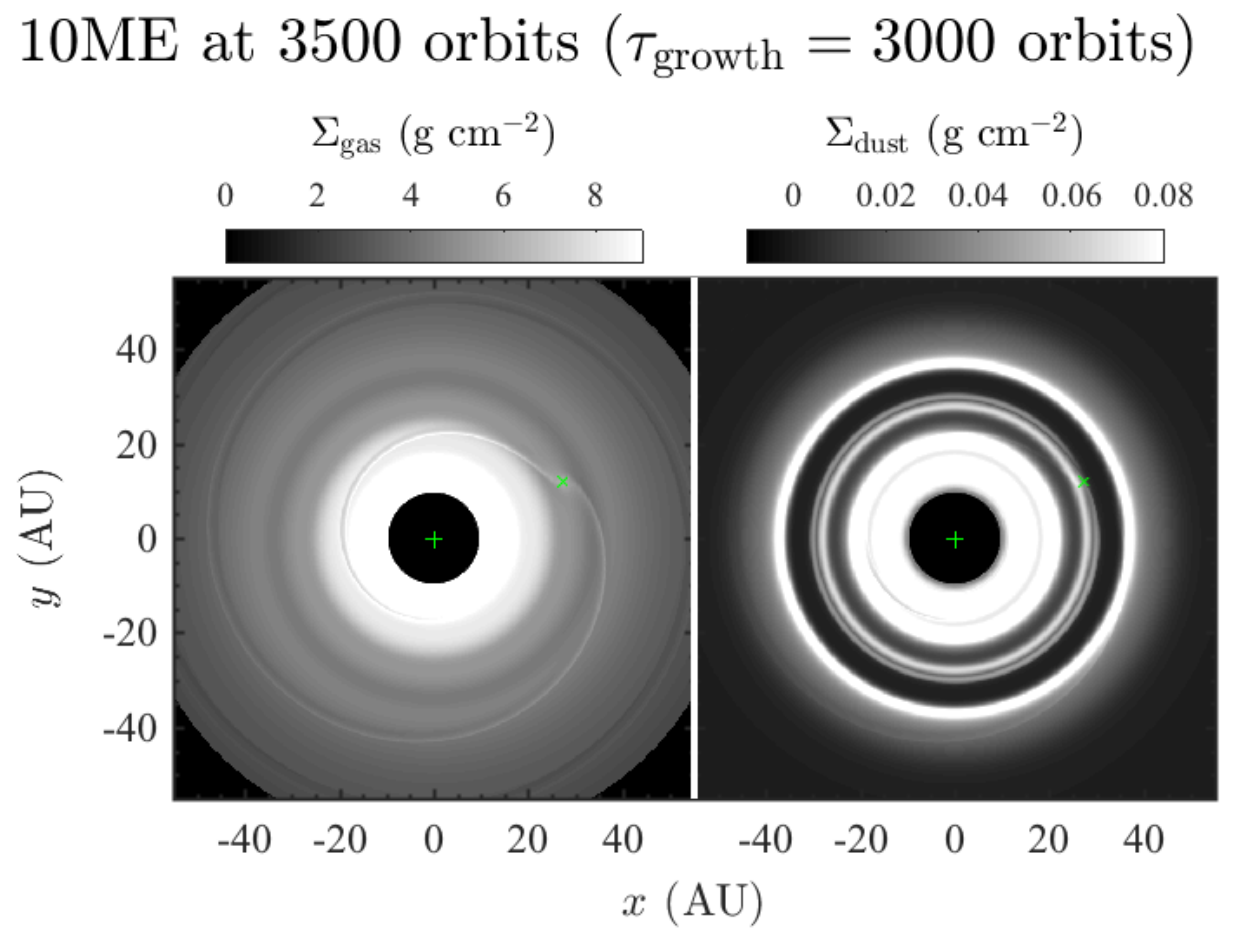}
\includegraphics[trim=0 0 0 0, clip,width=0.5\textwidth,angle=0]{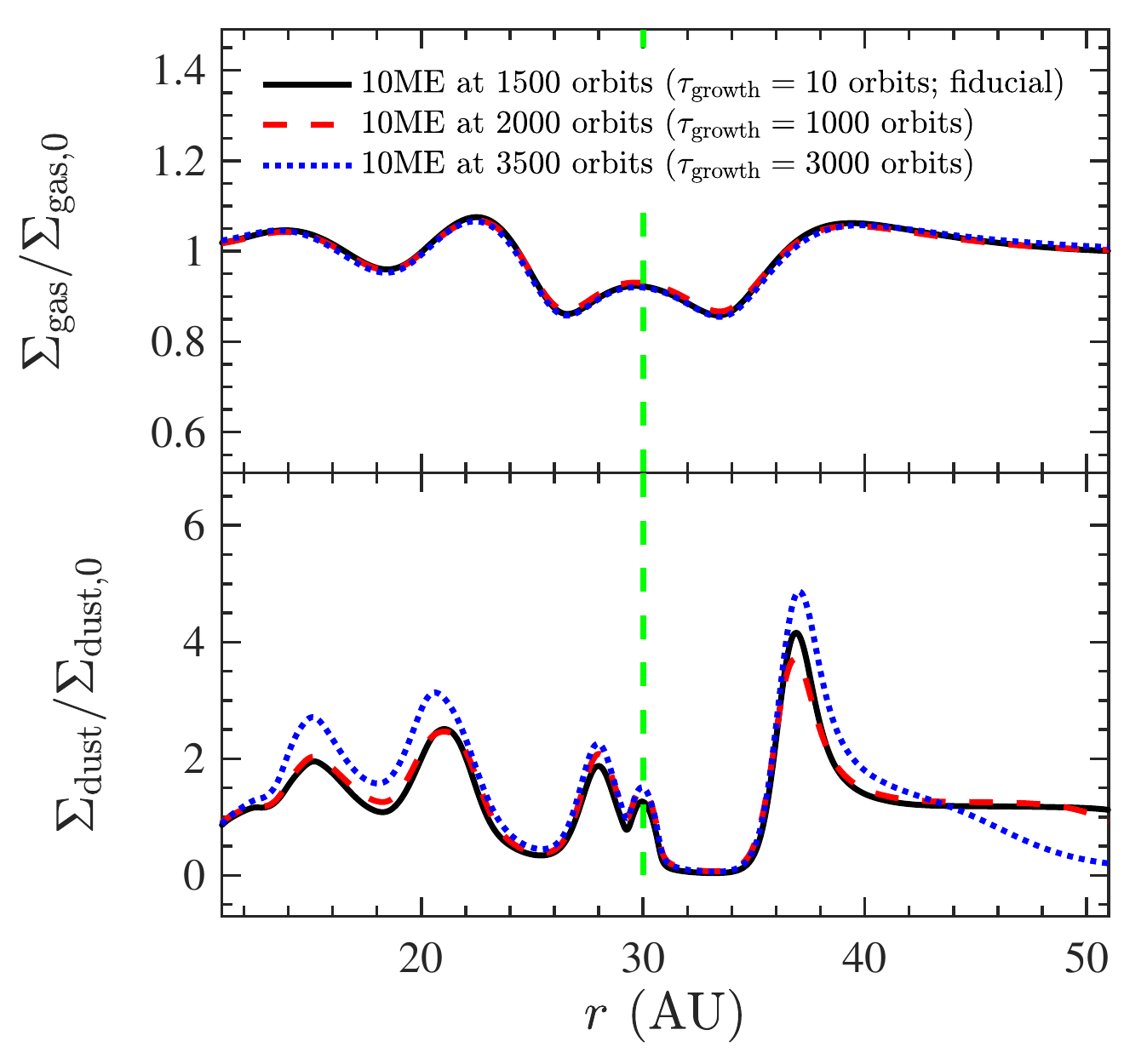}
\end{center}
\figcaption{{\bf Top}: Surface density maps of two models
having the same setup as our fiducial model 10ME, 
 but with $\mplanet$ ramped up over the first $\tau_{\rm growth} =1000$ orbits (left, sampled at $t=2000$ orbits) and $\tau_{\rm growth} = 3000$ orbits (right, sampled at $t=3500$ orbits), as compared to our standard $\tau_{\rm growth} = 10$ orbits. {\bf Bottom}: Azimuthally averaged radial profiles of the fiducial model 10ME and the two experimental models at 1500, 2000, and 3500 orbits, respectively. All three models take $\mplanet(t)\propto\sin{(t)}$ during the planet growth stage.
 These results show that the final appearances of gaps and rings are, not surprisingly, not affected by $\tau_{\rm growth}$. See Section~\ref{sec:10me-growth} for details.
\label{fig:sigma_rp_10me_growth}}
\end{figure}

In this section, we explore the effect of lengthening the planet growth timescale. Figure~\ref{fig:sigma_rp_10me_growth} compares our fiducial model 10ME ($\tau_{\rm growth}$=$10$ orbits) at 1500 orbits, with two other models having $\tau_{\rm growth} = 1000$ and 3000 orbits, sampled at 2000 and 3500 orbits, respectively.  The three models show similar surface density profiles in both gas and dust. Growing the planet in a few thousand orbits does not materially affect the multiple rings and gaps ultimately produced by the planet.\footnote{\citet{hammer16} found that more slowly growing planets trigger weaker vortices at gap edges.}

\subsection{Dependence on Planet Mass}\label{sec:mp}

\begin{figure}
\begin{center}
\includegraphics[trim=0 0 0 0, clip,width=\textwidth,angle=0]{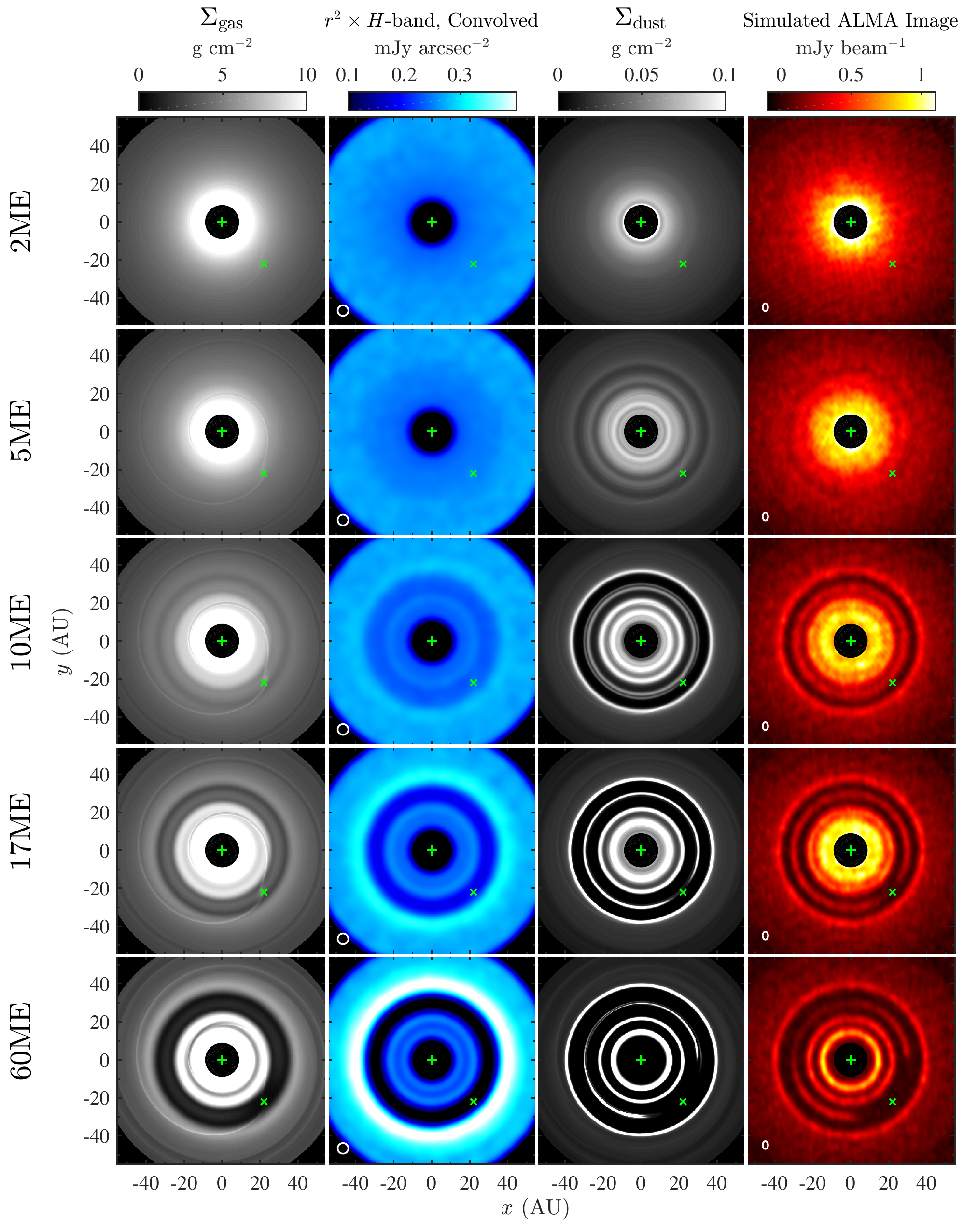}
\end{center}
\figcaption{Similar to Figure~\ref{fig:image_times}, but for different planet masses at 1500 orbits. The double-gap feature becomes more prominent with increasing $\mplanet$. At the highest planet masses considered, the middle ring develops azimuthal asymmetries. See Section~\ref{sec:mp} for details.
\label{fig:image_mp}}
\end{figure}

\begin{figure}
\begin{center}
\includegraphics[trim=0 0 0 0, clip,width=\textwidth,angle=0]{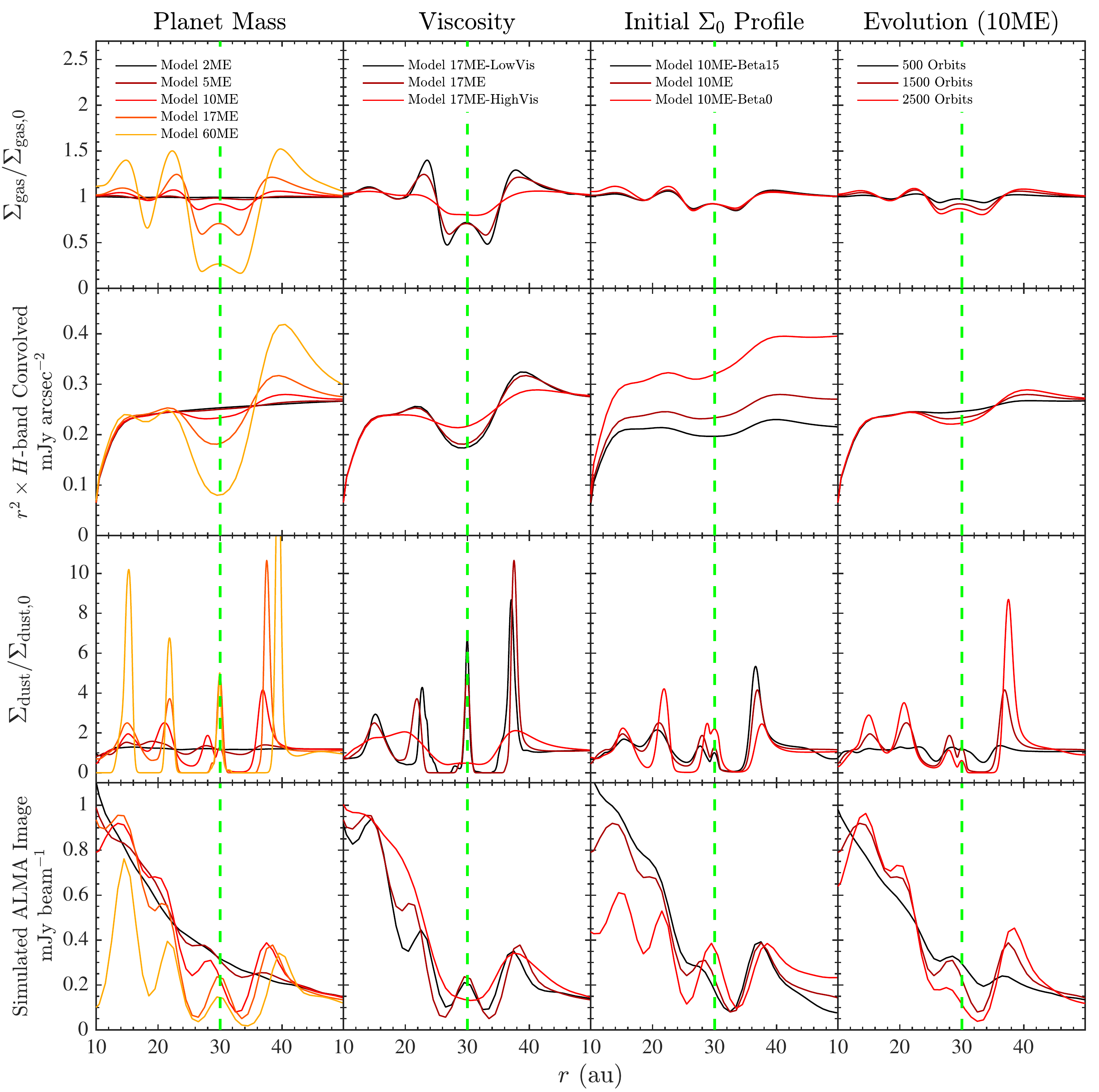}
\end{center}
\figcaption{{\bf From top to bottom}, azimuthally averaged radial profiles of $\sigmag/\Sigma_{\rm gas,0}$, $r^2$-scaled convolved $H$-band polarized
intensity, $\sigmad /\Sigma_{\rm dust,0}$, and simulated ALMA Band 7 flux density. {\bf From left to right}, models with different planet masses, viscosities, initial $\sigmag$ profiles, and timestamps. See Section~\ref{sec:results} for details.
\label{fig:rp}}
\end{figure}

\begin{figure}
\begin{center}
\includegraphics[trim=0 0 0 0, clip,width=0.5\textwidth,angle=0]{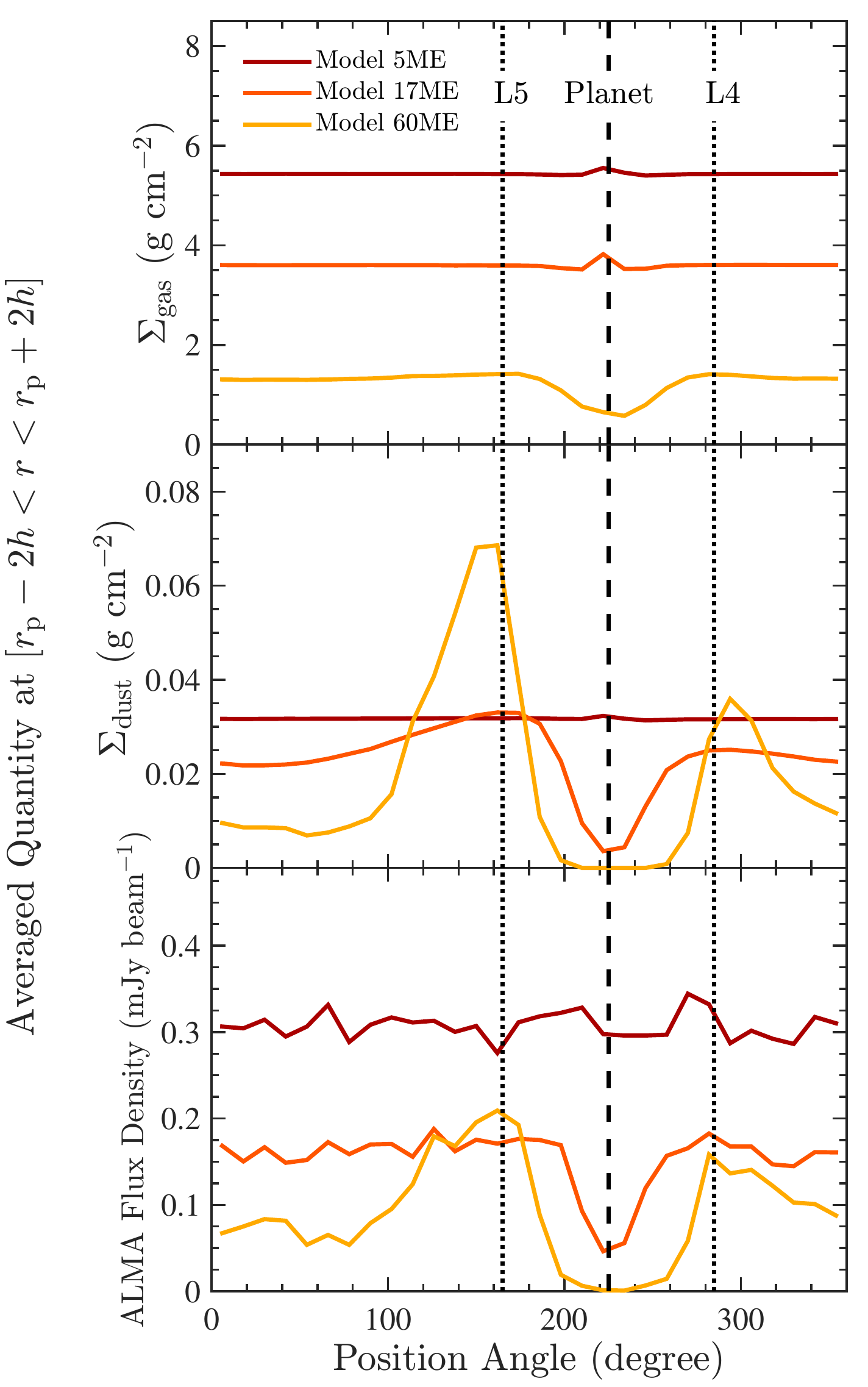}
\end{center}
\figcaption{{\bf From top to bottom}, azimuthal profiles, averaged radially over the co-orbital region, of the gas surface density, big-dust surface density, and the flux density in simulated ALMA images, for models with different planet masses (see also Figure~\ref{fig:image_mp}).
Quantities are averaged within the radial interval $\rp-2h<r<\rp+2h$ ($h=1.5$ AU at $r=\rp$). The position angles of the planet and of the two triangular Lagrange points L4 and L5 are labeled by the vertical dashed line and dotted lines, respectively. See Section~\ref{sec:mp} for details.
\label{fig:mp-azimuthal}}
\end{figure}

In this section, we explore how the morphologies of the rings and gaps vary with planet mass. Surface density maps and simulated images are shown in Figure~\ref{fig:image_mp}, radial profiles are shown in Figure~\ref{fig:rp} (first column), and azimuthal profiles of selected models are shown in Figure~\ref{fig:mp-azimuthal}. We find the following behavior when increasing $\mplanet$ from $2\me$ to $60\me$:
\begin{enumerate}
\item Gas gaps deepen with increasing $\mplanet$. As more gas is pushed to gap edges, $\sigmag$ in the IR and in the OR increases, leading to an increase in the gap-ring contrast in $H$-band images. In simulated ALMA images, the double-gap feature first appears at $\mplanet=5\me$ and grows more prominent with increasing $\mplanet$.
\item Each dust ring varies in its own way with $\mplanet$. For the MR, $\sigmag$ and by extension $\sigmad$ decrease with increasing $\mplanet$.  By comparison, $\sigmad$ in the HR rises. As a result, as $\mplanet$ increases in the simulated ALMA images, the location of the (practically merged) MR+HR shifts from slightly inward of the planet's orbit to being more nearly coincident with the planet's orbit. The IR, IR2, and OR all become more prominent in $\sigmag$ and $\sigmad$ as $\mplanet$ increases. In simulated ALMA images, IR and IR2 are distinguishable only for $\mplanet\gtrsim$ Neptune mass.
\item The OR is pushed outward.
\item As $\mplanet$ increases, the HR becomes more
azimuthally asymmetric. Figure~\ref{fig:mp-azimuthal} shows azimuthal
profiles of $\sigmag$, $\sigmad$, and ALMA flux density in the
coorbital region $r = \rp-2h$ to $r= \rp+2h$. In $\sigmag$, a 60 $\me$
planet starts to evacuate material in its vicinity to produce an
azimuthal dip (compare with Figure 9 of \citealt{yu10}). A similar
structure can also be found in the dust distribution and starts at a
smaller $\mplanet\sim$ Neptune mass, widening in angle with increasing
$\mplanet$. In addition, dust starts to accumulate around the
triangular Lagrange points L4 and L5 $\pm60^\circ$ away from the
planet, as seen in previous works \citep[see, e.g., the 40$\me$ planet
case in][Figure 5]{zhu14votices}. The simulated ALMA image with $\mplanet=60\me$ shows two Trojan arcs (bottom right panel in Figure~\ref{fig:image_mp}). Eventually, as $\mplanet$ approaches a Jupiter mass \citep{lyra09}, the dust HR retreats into two azimuthally confined spots around L4 and L5, and the double-gap feature vanishes.
\end{enumerate}

In general, dust gaps are more prominent than gas gaps, consistent with \citet{paardekooper06}.
We note that the minimum planet mass for the double gap to appear in observations ($5\me$ in our models) depends on disk parameters; for example, the minimum $\mplanet$ decreases with decreasing scale height (data not shown).

\subsection{Dependence on Viscosity}\label{sec:viscosity}

\begin{figure}
\begin{center}
\includegraphics[trim=0 0 0 0, clip,width=\textwidth,angle=0]{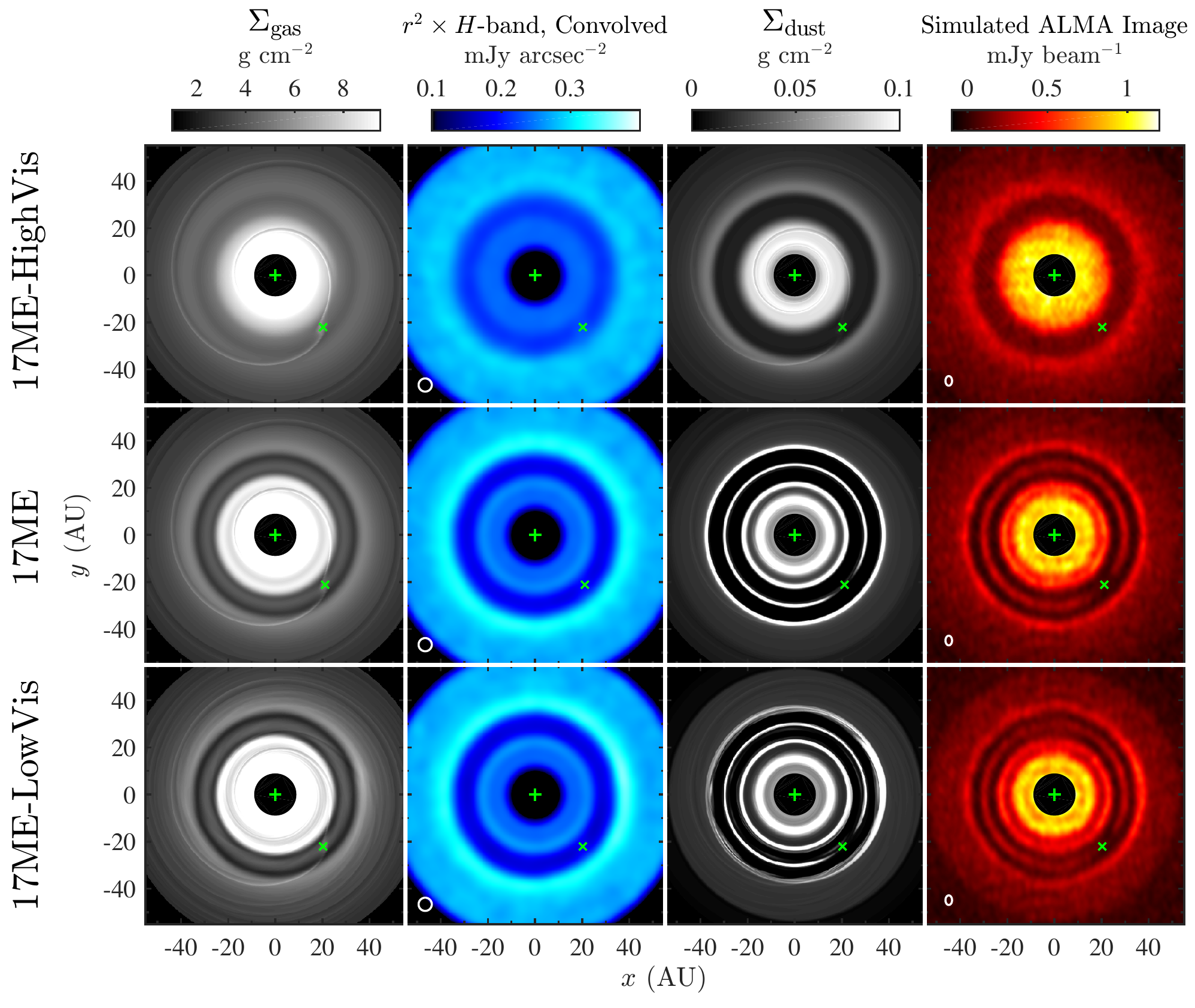}
\end{center}
\figcaption{Similar to Figure~\ref{fig:image_mp}, but for different viscosities. A value for $\alpha$ on the order of $10^{-3}$ destroys
the double-gap feature. See Section~\ref{sec:viscosity} for details.
\label{fig:image_viscosity}}
\end{figure}

Viscous diffusion tends to smooth out density perturbations in the gas
induced by planets. In Figures~\ref{fig:rp} (second column) and 
\ref{fig:image_viscosity}, we compare three simulations with a
Neptune-mass planet and different values for $\alpha$: 5$\times$10$^{-4}$, 5$\times$10$^{-5}$ (our fiducial model of Section~\ref{sec:10me}), and 5$\times$10$^{-6}$. 

In the $\sigmag$ snapshots and the NIR images, as viscosity decreases
the gaps deepen, as expected \citep[e.g.,][]{fung14, kanagawa15,
  duffell15gap}. A second concomitant effect is that the surface
densities in the IR and OR, located at the gap edges, increase with
decreasing viscosity: when $\alpha$ drops, more gas is pushed by
Lindblad torques into the gap edges, and low viscosity makes it harder
to diffuse away this excess. A third effect is that at the highest
viscosity ($\alpha=5\times10^{-4}$), the HR+MR in the gas disappears and
the two gaps merge. High viscosity enables gas to more easily
diffuse away from the co-orbital region, which is a potential maximum.
In addition, for planets with $\mplanet\lesssim\mth$, viscous diffusion can prevent waves from shocking and damping and can thereby suppress gap formation \citep{goodman01, dong11nonlinear}.
All three effects have been seen in previous hydro simulations \citep[e.g.,][]{li09, yu10}.

\begin{figure}
\begin{center}
\includegraphics[trim=0 0 0 0, clip,width=\textwidth,angle=0]{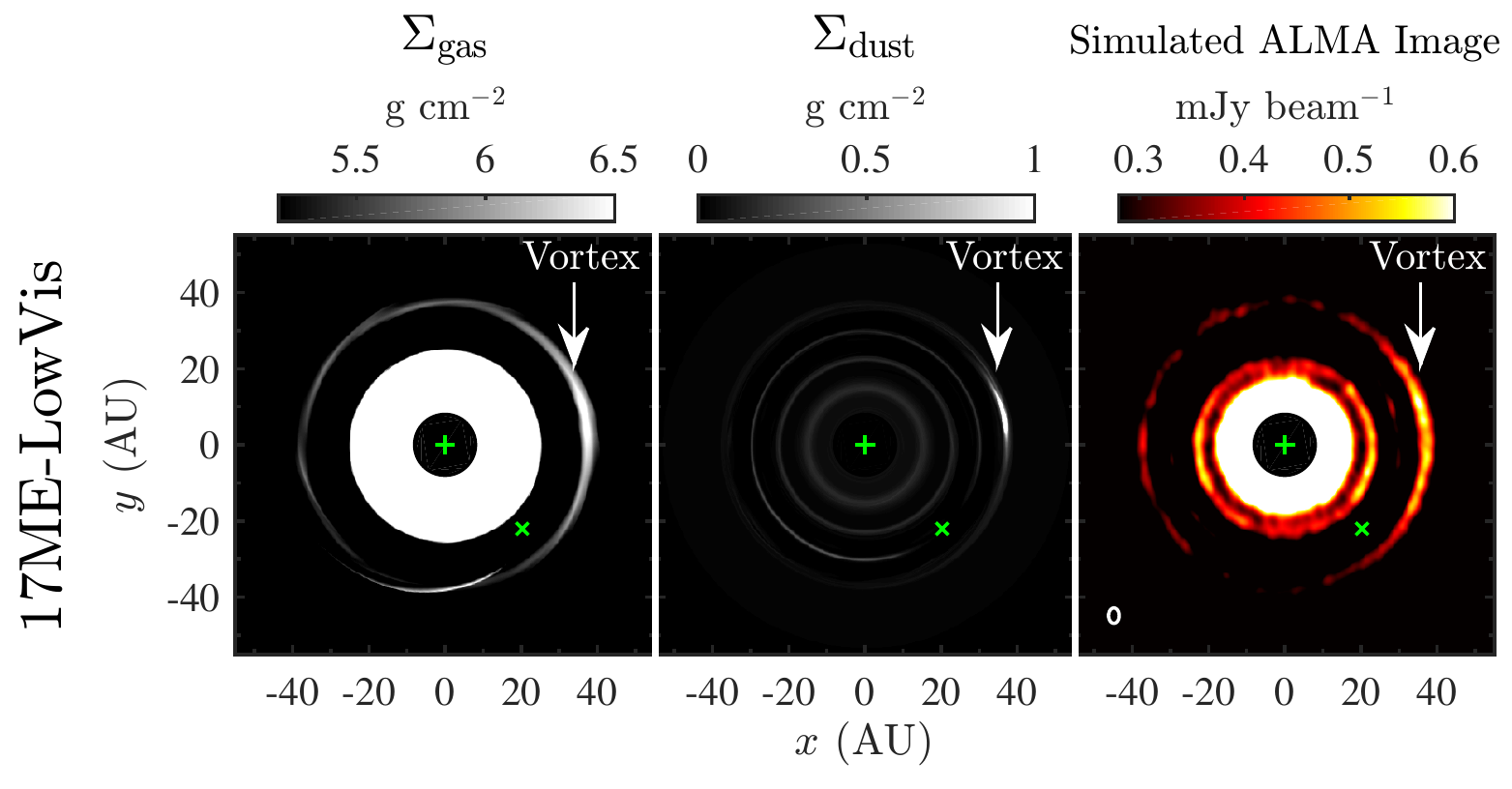}
\end{center}
\figcaption{Model 17ME-LowVis (shown in the bottom row of Figure~\ref{fig:image_viscosity}) with more dramatic color stretches to highlight the vortex at the gap edge. See Section~\ref{sec:viscosity} for details.
\label{fig:17me_lowvis}}
\end{figure}

In the $\Sigma_{\rm dust}$ snapshots and simulated ALMA observations,
the two low viscosity runs produce
qualitatively similar ring and gap structures. By contrast, in the
$\alpha=5\times10^{-4}$ run, the two gaps merge and only two broad
overdensities remain at the gap edges. A wide single gap may not be distinguishable from the gap opened by a more massive planet in a disk with higher viscosity. 
Thus, low viscosity ($\alpha < 10^{-3}$) is a necessary
condition for super-Earths to induce their signature double gaps. The
two low viscosity models differ, however, in that for
$\alpha=5\times10^{-6}$, a vortex is created,
as indicated in Figure~\ref{fig:17me_lowvis} using a more dramatic color
stretch.
The development of vortices at edges of planet-opened gaps requires
low viscosity \citep[e.g.,][]{zhu14stone}, and is probably
driven by the RWI \citep[e.g.,][]{li00, li01}.
As with dust trapping by radial pressure structures,  pressure peaks in
azimuth associated with vortices also trap dust particles of certain sizes \citep[e.g.,][]{lyra13}. The vortex in 17ME-LowVis is quite visible in the simulated ALMA image (Figure~\ref{fig:17me_lowvis}, right panel).

\subsection{Dependence on the Global Background Disk Profile}\label{sec:sigma0}

\begin{figure}
\begin{center}
\includegraphics[trim=0 0 0 0, clip,width=\textwidth,angle=0]{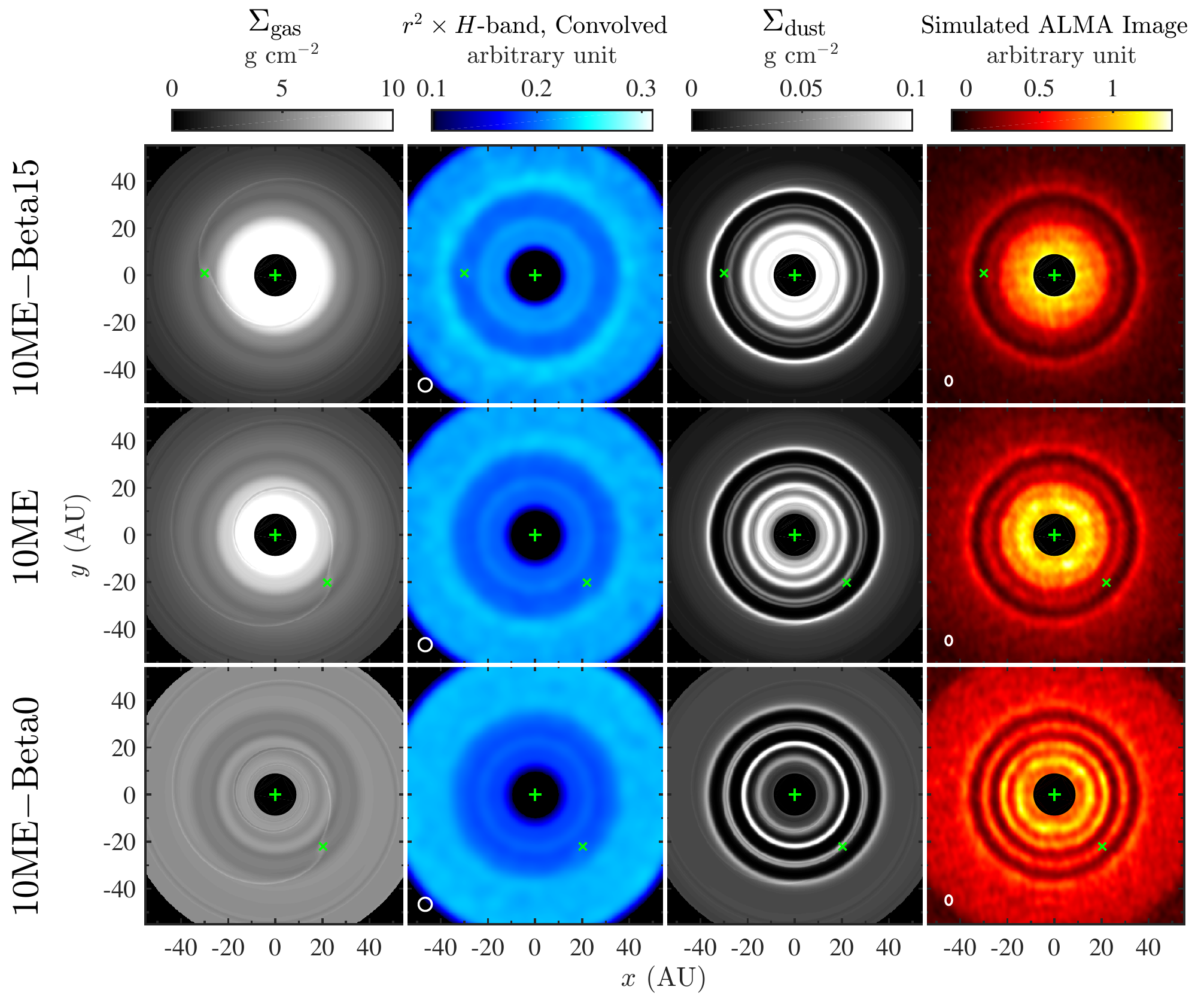}
\end{center}
\figcaption{Similar to Figure~\ref{fig:image_mp}, but for different initial surface density profiles. The locations and contrasts of the rings and gaps are affected by the choice of profile. See Section~\ref{sec:sigma0} for details.
\label{fig:image_sigma0}}
\end{figure}

The dust distribution, a direct observable from ALMA, is determined by
gas drag forces, which depend on local gas pressure gradients.
These local gradients in turn depend on the 
global background gradient.
Figures~\ref{fig:rp} (third column) and \ref{fig:image_sigma0}
compare models 10ME-Beta15 ($\Sigma_0\propto1/r^{1.5}$), 10ME
($\Sigma_0\propto1/r$), and 10ME-Beta0 ($\Sigma_0\propto$
constant). The three models share similar radial profiles for
$\sigmag/\Sigma_{\rm gas,0}$, but the locations and contrasts of the
dust rings depend on $\beta$, the power-law slope of the
background surface density.
As $\beta$ decreases, the IR, HR, MR, and OR in $\sigmad/\Sigma_{\rm
  dust,0}$ expand outward.
Overall, in the simulated ALMA observation,
the 4 rings in model 10ME-Beta0 appear to be more distinctive
and well separated than in the other two models. 

\subsection{Effects of Planet Migration}\label{sec:migration}

Finally, we experiment with the planet's radial migration. 
We manually move the planet in models 10ME-Beta0 and 17ME across the disk at a constant
user-controlled migration rate to explore how planet-induced
structures depend on migration. 

\begin{figure}
\begin{center}
\includegraphics[trim=0 0 0 0, clip,width=\textwidth,angle=0]{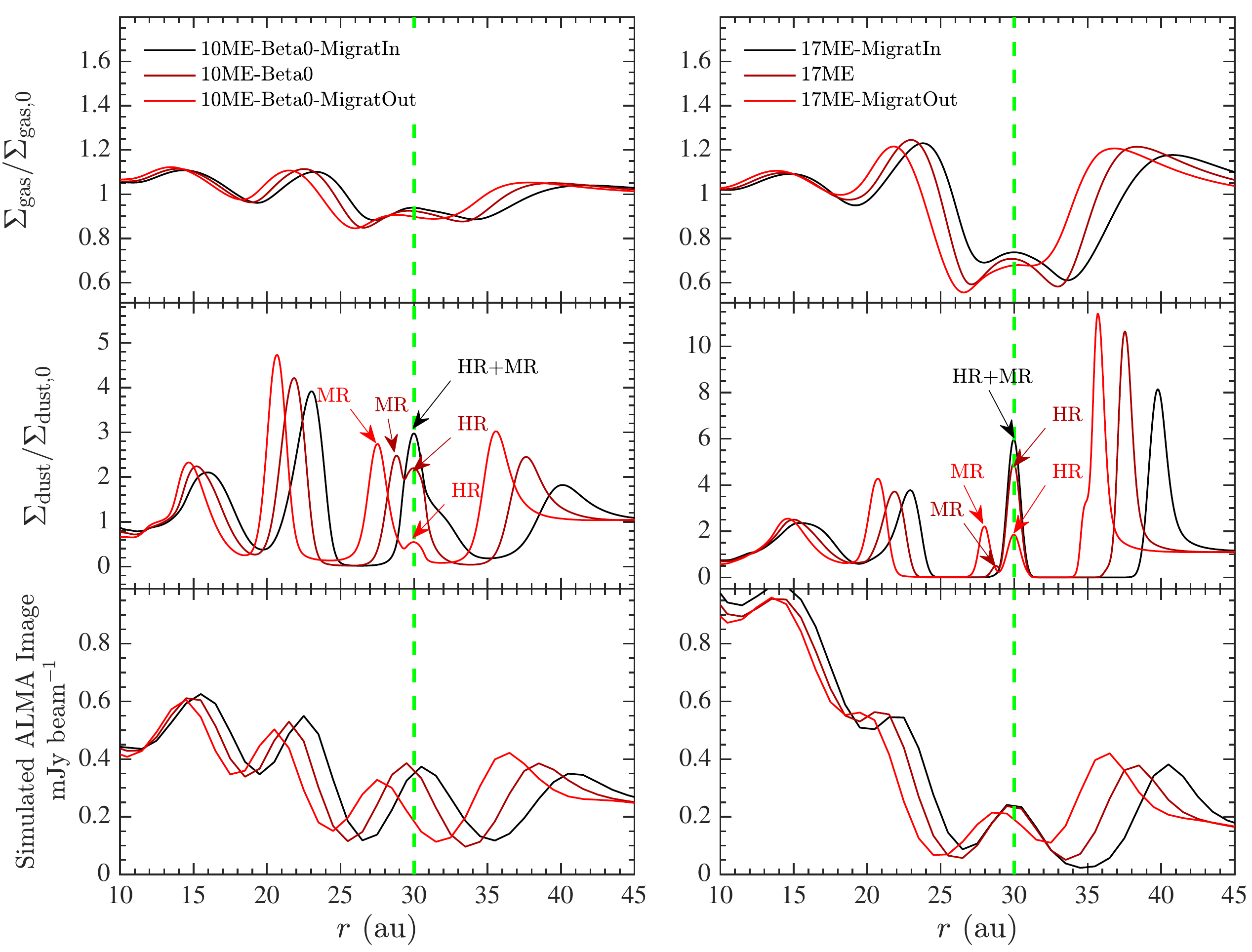}
\end{center}
\figcaption{Similar to Figure~\ref{fig:rp}, but for migrating planets. The double-gap feature persists under a modest migration rate ($|r/\dot{r}|\sim$ a few Myr). The MR and HR in the dust density panels are labeled. When the planet is migrating (either inward or outward),
those features radially behind the planet 
lag further behind, while features radially ahead of the planet 
are squeezed closer to the planet. The one exception is the dust HR,
which follows the planet. See Section~\ref{sec:migration} for details.
\label{fig:rp_migration}}
\end{figure}

\begin{figure}
\begin{center}
\includegraphics[trim=0 0 0 0, clip,width=\textwidth,angle=0]{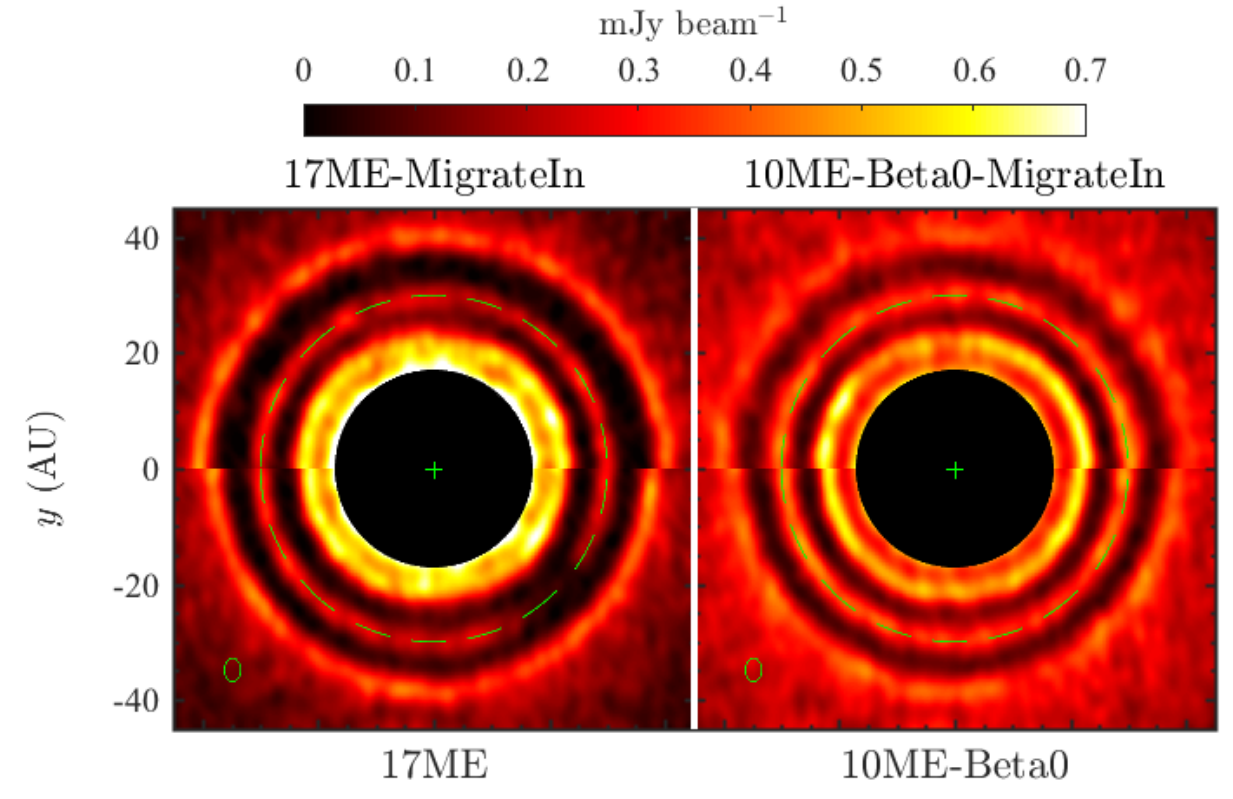}
\end{center}
\figcaption{Simulated ALMA images showing the effects of planet migration. The inner 17 AU is masked out to highlight the outer disk. Each panel contains two models separated along the horizontal central line. The planet's orbit is indicated by the dashed green circle ($\rp=30$ AU in all models). In the 17ME models ({\bf left}), the MR+HR is dominated by the HR, which simply follows the planet, while all other rings in 17ME-MigrateIn shift outward. Thus, the inner gap in 17ME-MigrateIn is compressed while the outer gap is widened. In the 10ME-Beta0 models ({\bf right}), the MR+HR is dominated by the MR; all features in the MigrateIn model shift outward, with the gap widths staying roughly constant. The radial profiles of these models are shown in Figure~\ref{fig:rp_migration}. See Section~\ref{sec:migration} for details.
\label{fig:image_migration}}
\end{figure}

Figures~\ref{fig:rp_migration} and
\ref{fig:image_migration} show
the results. In models 10ME-Beta0-MigrateIn and 17ME-MigrateIn, the planet migrates
inward at a rate $\dot{r}=-2\times10^{-3}$AU orbit$^{-1}$ (or about
$-10^{-5}$AU year$^{-1}$). In models 10ME-Beta0-MigrateOut and 17ME-MigrateOut, the
planet migrates at the same rate but outward. In all four models,
the planet 
crosses $\rp=30$ AU at the end of the simulation after 1500 orbits. 
We note that it takes 1500 orbits for the planet to migrate 3 AU (= 2 scale heights), about the distance between the nearest gap and the planet's orbit. 
We find that when the planet migrates (either inward or outward),
those features radially behind the planet 
lag further behind, while features radially ahead of the planet 
are squeezed closer. The one exception is the dust HR,
which simply follows the planet.
The radial shifts of the IR2, IR, MR, and OR
could, in principle, be used to infer if a planet
is migrating.
For example, if the HR+MR ring in ALMA images is dominated by the HR, the gap
ahead of a migrating planet narrows, while the gap behind the migrating
planet widens, as shown in the left panel of
Figure~\ref{fig:image_migration}.
However, reading the signature of migration
would require knowledge of the background
disk profile to determine the ``intrinsic'' locations of these
features in the absence of migration.

\subsection{Future Improvements}\label{sec:future}

Our investigations of observational signatures of super-Earths
can be
extended for greater realism.
\begin{enumerate}
\item 
Simulations with a more realistic dust size distribution are a natural 
extension \citep[e.g.][]{ruge16}.
In this work we chose 200 $\micron$ as a representative size for the ``big''
dust particles. But dust of other sizes
can have a non-trivial opacity at $\sim$mm wavelengths.
\item Relatedly, grain growth and evolution can be included. Once dust is
concentrated into rings, coagulation can be
facilitated \citep[e.g.,][]{youdin02, johansen07, pinilla16}, and
instabilities such as the streaming instability \citep{youdin05, chiang10}
can be triggered. Our simulations focus on the first stage, the drift of
$\sim$mm-sized dust into gas pressure perturbations.
A natural extension would be to follow the subsequent evolution of dust in
the resulting concentrations. 
\item Our simulations assume a non-zero $\alpha$ viscosity. Truly inviscid disks can behave differently \citep{fung17}.
\item Our simulations are 2D. For planets with $\mplanet>\mth$, \citet{fung16} have shown that gap opening proceeds in 3D much the same way that it does in 2D. It will be interesting to explore 3D effects of sub-thermal-mass planets in low viscosity disks. 
\end{enumerate}


\section{Summary and Discussion}\label{sec:summary}

In this work, we combined two-fluid hydrodynamics and radiative transfer
simulations to calculate the distributions of gas and of
$\sim$mm-sized dust in a low viscosity disk perturbed by a super-Earth, and
created synthetic observations of such systems in both near-infrared (NIR)
scattered light and ALMA mm dust continuum emission. Our main findings are:
\begin{enumerate}
\item A $\sim$10 $\me$ planet in a low viscosity disk initially opens two narrow
annular gaps, each displaced radially by about a
vertical pressure scale height to either side of the planet's orbit. Perturbations in the gas pressure cause dust particles to drift differentially. Millimeter-sized dust particles concentrate into three distinctive rings: two rings at the far edges of the two gaps, and one ring sandwiched between the two gaps (Figure~\ref{fig:sigma_10me}). The 1:1 resonance may produce a fourth dust ring exactly coincident with the planet's orbit. A fifth dust ring interior to the double-gap region is also possible, depending on parameters. These dust rings are not necessarily located at the positions of low-order mean motion resonances relative to each other, although they may appear to be so by chance. For our assumed disk parameters, a minimum planet mass of $\sim$5$\me$ is needed to produce detectable rings on $\sim$1000 orbits timescale. Planets more massive than Neptune ($\gtrsim 17 \me$)
generate azimuthal dust concentrations at the triangular Lagrange points L4 and L5.
\item The dust rings and gaps produced by a super-Earth at 30 AU in a typical disk with $\sim$$10$~$\me$ of mm-sized dust at 140 pc distance can be readily detected by ALMA in mm dust continuum observations with an angular resolution of $0\arcsec.03$ and 2 hours of integration. In NIR scattered light, the gaps and rings are less prominent and may not be detectable with current NIR capabilities.
\item To produce the signature double-(or-multiple)-gap feature in mm continuum observations, the disk's $\alpha$ viscosity parameter has to be lower than $\sim$10$^{-4}$. At higher viscosity, only one gap is opened at the planet's orbit. If $\alpha$ is lower than $\sim$10$^{-5}$, the Rossby wave instability at gap edges can be triggered, forming dust-trapping vortices.
\item The double-gap signature of a single super-Earth
manifests in disks with a variety of background profiles 
and tolerates
modest planet migration (10 AU / Myr at 30 AU).
Migration changes the locations of the rings in potentially observable ways.
Although the double gap can be discerned for 
thousands of orbits, it weakens
as material diffuses away from the co-orbital region.

\end{enumerate}

At tens of AUs, super-Earths can produce a characteristic double-gap
feature detectable in high resolution mm continuum observations by
ALMA. 
A low disk viscosity of $\alpha\lesssim10^{-4}$ is critical
for observability. Super-Earths, unlike their Jovian siblings,
induce only weak perturbations in the gas disk, which must be
nearly inviscid for such perturbations to manifest observationally
(in particular, to facilitate the development of the
\citealt{goodman01} wave nonlinearity), and for aerodynamic
drag on dust to amplify these features.
The low viscosity at the disk midplane that is
advocated here finds support from both 
observations and theory. The leading explanation for lopsided
disks observed by ALMA \citep[e.g.,][]{vandermarel13, casassus13,
  perez14}---vortices formed at planet-opened gap edges---requires
$\alpha\lesssim10^{-4}$ \citep[e.g.,][]{zhu14stone}. Similarly, if
most cavities in transitional disks are opened by multiple planets,
$\alpha$ needs to be lower than $10^{-3}$ in order to mobilize enough
low-mass planets to account for the
prevalence of transitional disks \citep{dong16td}.
See also \citet{fung17} for how multiple super-Earths in inviscid disks 
can both survive orbital migration and drive disk accretion.
Magnetohydrodynamics (MHD) simulations have shown than
non-ideal MHD effects, in particular ambipolar diffusion, can
significantly suppress the magnetorotational instability in disks at
tens of AUs, resulting in practically zero turbulence at disk midplanes
\citep{bai11stone,perezbecker11-td, perezbecker11,
  bai13, turner14, bai15}.
Not even in disk surface layers has
turbulence been detected (\citealt{flaherty15}; Flaherty et al.~2017, in prep.).

A double gap observed in nature could, naively, be explained
with two planets in a standard viscous disk, each opening their own 
gap. But this interpretation
is problematic. The two gaps in the double-gap
feature that we found are typically $\sim$2--4 scale heights apart, or 0.1--0.2
$\rp$ for $h/r=0.05$. In a disk with $\alpha>10^{-4}$, the width of a
single planet-opened gap is greater than $0.2\rp$ (e.g., \citealt{kanagawa16width}, Figure 3; \citealt{dong17gap}, Figure 7). Two
closely spaced gaps opened by two planets would merge into one
in a standard viscous disk.
The two-planet configuration may also be prone to orbital instability, particularly if the planets are not in mean-motion resonance \citep{tamayo15}. 

We close with a few remarks on real disks. To date, two systems,
HL Tau \citep{brogan15} and TW Hya \citep{andrews16,
  tsukagoshi16}, have been observed by ALMA with sub-$0\arcsec.05$
resolution.
Intriguingly, in each case a candidate double gap can be
identified: D5 and D6 at 64 and 74 AU in HL Tau, and the two gaps
at 37 and 43 AU in TW Hya. We plan to discuss these two systems and
explore the possibility that their ring/gap
features are produced by super-Earths
in a forthcoming paper. In addition, a Large Program (PI:
S. Andrews) was approved in ALMA Cycle 4 to resolve 20 nearby disks
at high angular resolution.
We look forward to seeing whether this sample also features
multiple rings, and ultimately to inferring
whether super-Earths commonly form at
large distances, as they appear to have done at small ones.


\section*{Acknowledgments}

We are grateful to an anonymous referee for constructive suggestions that improved our paper. R.D. thanks Bekki Dawson, Jeffrey Fung, Eve Lee, Paola Pinilla, Roman Rafikov, and
Zhaohuan Zhu for insightful discussions; Fumi Egusa at the ALMA help
desk on using CASA; and Sean Andrews and David Wilner for motivating
this work. Part of the numerical calculations were performed on the
SAVIO cluster provided by the Berkeley Research Computing program,
supported by the UC Berkeley Vice Chancellor for Research and the
Berkeley Center for Integrative Planetary Science. H. Li and S. Li
gratefully acknowledge the support by LANL's LDRD program and a CSES
project. E.C. acknowledges support from the NSF and NASA.


\clearpage
\begin{table}[]
\footnotesize
\centering
\begin{tabular}{c|ccccc}
\hline
Model Name & $\mplanet$ & $\alpha$ & Initial $\Sigma_0-r$ & Migration \\ \hline
2ME  & $2\me$ & $5\times10^{-5}$ & $\propto 1/r$ & No \\ 
5ME   & $5\me$ & $5\times10^{-5}$ & $\propto 1/r$ & No    \\ 
10ME   & $10\me$ & $5\times10^{-5}$ & $\propto 1/r$ & No    \\ 
17ME   & $17\me$ (Neptune) & $5\times10^{-5}$ & $\propto 1/r$ & No    \\ 
60ME   & $60\me$ & $5\times10^{-5}$ & $\propto 1/r$ & No   \\ 
\hline
10ME-Beta0  & $10\me$ & $5\times10^{-5}$ & Constant  & No  \\ 
10ME-Beta15  & $10\me$ & $5\times10^{-5}$ & $\propto 1/r^{1.5}$  & No  \\ 
\hline
17ME-LowVis  & $17\me$ & $5\times10^{-6}$ & $\propto 1/r$ & No  \\ 
17ME-HighVis  & $17\me$ & $5\times10^{-4}$ & $\propto 1/r$ & No  \\ 
\hline
10ME-Beta0-MigrateIn  & $10\me$ & $5\times10^{-5}$ & Constant & $\dot{r}=-2\times10^{-3}$AU orbit$^{-1}$  \\ 
10ME-Beta0-MigrateOut  & $10\me$ & $5\times10^{-5}$ & Constant & $\dot{r}=2\times10^{-3}$AU orbit$^{-1}$  \\ 
17ME-MigrateIn  & $17\me$ & $5\times10^{-5}$ & $\propto 1/r$ & $\dot{r}=-2\times10^{-3}$AU orbit$^{-1}$  \\ 
17ME-MigrateOut  & $17\me$ & $5\times10^{-5}$ & $\propto 1/r$ & $\dot{r}=2\times10^{-3}$AU orbit$^{-1}$  \\ 
\hline
\end{tabular}
\caption{Models. In all non-migration simulations, the planet's orbit is fixed
at $\rp=30$ AU, where $h/r=0.05$ ($\mth=40\me$).
In the last four models with a migrating planet, the migration rate is fixed at
$|\dot{r}|=2\times10^{-3}$AU orbit$^{-1}$ (or about
$10^{-5}$AU year$^{-1}$); the planet reaches $\rp=$ 30 AU after 0.25 Myr.}
\label{tab:models}
\end{table}

\end{document}